# Observer-Based Consensus of Nonlinear Positive Multi-Agent Systems with Saturated Control Input*

Amirreza Zaman, Wolfgang Birk, Khalid Tourkey Atta

*Control Engineering Group, Department of Computer Science, Electrical and Space Engineering, Luleå University of Technology, Luleå, Sweden.*

**Abstract**

This paper presents the distributed pinning consensus solution for nonlinear positive multi-agent systems with nonlinear control input by applying observer-based control protocols. The network topology is considered as a directed and fully connected structure. By considering sector input nonlinearities and various forms of topologies, two kinds of state observers involving standard observer and distributed pinning observer are presented for each regarded nonlinear agent by applying a novel analysis directly dealing with the nonlinear input and nonlinear system dynamics. The measured local output detail outlines the first observer, and the other observer is achieved via the corresponding output detail of its adjacent agents. Based on further observed state details, a distributed pinning observer-based strategy is derived for the leader-follower non-negative global consensus of the nonlinear positive multi-agent system. Additionally, two multi-step algorithms are proposed to set up the observer gains and each protocol criterion. Performance evaluations are provided to confirm the proposed control method and illustrate the effectiveness of the derived non-negative consensus observer-based protocols.

*Keywords:* Convex optimization; directed topology; distributed pinning observer; linear matrix inequality(LMI); nonlinear control input; nonlinear positive multi-agent systems.

## 1. Introduction

The distributed cooperative control approach of multi-agent systems (MASs) has been investigated previously in many cases. These systems are applicable in control, sensing, communication network technologies, and district heating and cooling networks. Other substantial perspectives of analyzing distributed MASs are using consensus or synchronization control problems in recent years [1] since most of the coordination control problems can be stated into consensus problems [2]. Besides, by considering the physical features of technical systems such as nonlinearity, actuation, and model uncertainties, the latest consensus studies have been developed based on these physical factors [3, 4, 5, 6] to improve the networks' performance.

Most recently proposed consensus protocols considered general MASs with undirected networks [7, 8]. Since the Laplacian matrices of MASs are asymmetrical, analyzing consensus problems of nonlinear MASs in directed networks becomes much more complicated than undirected approaches. Thus, some distinctive approaches have been made to design fully distributed consensus protocols for directed topologies [9, 10]

*Amirreza Zaman(Corresponding author)
⋆ Funding received from the Horizon 2020 Research Programme of the European Commission under the grant number 956059 (ECO-Qube) is hereby gratefully acknowledged.
*Email addresses:* amirreza.zaman@ltu.se (Amirreza Zaman), wolfgang.birk@ltu.se (Wolfgang Birk), khalid.atta@ltu.se (Khalid Tourkey Atta)

because distributed networks can be more flexible and computationally efficient according to their sparse communication networks. Distributed consensus studies are categorized into regulator-based consensus problems [11, 12] and the tracking-based consensus problems [13, 14, 15, 16]. The majority of the studies are related to analyzing tracking consensus problems with single or more leaders in many areas, for instance, in power grids with various distributed generation units [17, 18]. The main issue in developing distributed tracking consensus cases is sharing objective information with agents. To overcome this matter, considering that agents can receive information from the desired trajectory, the consensus tracking problem can be dealt with using classical centralized methods. In some cases when defining consensus strategies, it is desired that all the states converge to a prescribed value. For this purpose, one can make each state converge to the prescribed value by employing a negative feedback term applied to each agent. Interestingly, it is not mandatory to have controllers on all nodes due to the network's interaction. This technique is known as pinning control. For extensive networks with several agents [19], it is not affordable and easy to compute agents one by one to achieve the desired consensus information. In such networks, a distributed pinning tracking control method is the solution to obtain the desired trajectory by sharing that information with a small number of agents. However, it is much more of interest for control engineers in providing pinning-based control strategies by sharing information with only one agent, which makes the implementation of the presented method much more straightforward.

Later, consensus strategies are obtained according to observer-based methods as some of the states may be unavailable in a multi-agent network [20, 21, 22]. In [23], distributed observers are applied to estimate the velocity of an assumed leader to reach and provide tracking/formation control of the network. Additionally, the distributed observer-based protocol based on relative output information is investigated in [24]. Moreover, a reduced-order observer-based consensus solution is proposed in [25]. Developing distributed consensus protocols by using edge/node information or utilizing intermittent observers are considered in [26] and [27].

For reduced-order observer-based consensus, an altered algorithm is proposed in [28] by bringing up the system's stability measure using the average dwell time approach and piecewise Lyapunov functions. Because the final goal is to implement multi-agent strategies in reality, it is obligatory to consider nonlinearities, saturations, and uncertainties in mathematical modelings to develop optimal consensus protocols [29, 30, 31, 32, 33]. Furthermore, to provide local consensus conditions for nonlinear MASs, the effects of the minimum nonzero eigenvalue of the Laplacian matrix with general algebraic connectivity consideration and spanning trees in directed topologies are analyzed in [34]. First and second-order MASs with intrinsic Lipschitz nonlinearities are investigated in [35] with fixed or switching directed networks to exploit the influence of the interaction concepts of positions and velocities in these networks. As a more advanced work about applying Lipschitz nonlinearities, an adaptive fully distributed consensus method without employing the global interaction information is introduced in [36]. For the goal of synchronizing linear undirected multi-agent networks, a control framework is developed in [37] by considering sector nonlinearities. The cyclic-small gain approach is introduced in [38] to obtain observer-based consensus under bounded uncertainties to lead agents to a limited region of the favorite consensus information. Also, the asymptotic and finite-time consensus of MASs under actuator saturation with or without time delays is studied in [39]. The low gain feedback method is considered, and generalized to an observer-based consensus scheme in [40] and [41] to provide semi-global consensus under input saturations in switching networks. Moreover, for undirected and directed multi-agent networks, an observer-based leader-following consensus for agents with uncertain dynamics is proposed in [42].

Positive systems are branches of systems in which their state variables and outputs remain non-negative for non-negative initial conditions and inputs. These systems have been studied more due to their applications in various essential areas, such as economics, biology, communications, population models, heat exchangers, distillation columns, storage systems, and stochastic models with non-negative states, *etc*[43, 44]. For instance, even the R-C circuit can represent a positive system because the non-negative value of the capacitor's voltage is initially such [44]. Since positive systems are often occurring in real-life contexts, it is necessary to study their properties. Some positive systems' applications are discussed in [44] in complete details, such as Leontief models for price and production's predicting, the Leslie model to study age-structured populations, the Markov chains, the compartmental models, and the birth and death processes. Additionally, positive interconnected systems are analyzed from the stability viewpoint by [45] with the $L_1$ gain test. The authors specified acceptability and stability measures of positive systems with the L1 induced norm of each positive subsystem and with the calculated Frobenius eigenvalue of the determined interconnection matrix [46]. For the goal of analyzing the exponential stability of switched positive linear systems with time-varying delays, [47] presented a weak excitation. Further, new outcomes on the stability of switched positive systems under time delays by considering copositive Lyapunov function are presented in [48]. Another approach [49] introduced the stability conditions of discrete-time positive systems with some conditions in their inputs or states.

Positive MASs are introduced when the state variables in MASs remain non-negative. Recently, positive MASs have been remarkably gained attention to the study. As an example, in [50], the containment control problem of MASs is solved by applying positive systems' fundamentals under heterogeneous and unbounded communication delays. Furthermore, the positive edge-consensus approach for nodal networks is developed in [20] by employing output feedback schemes. Discrete-time positive edge-consensus for both directed and undirected nodal networks is studied in [51]. For complex networks, the positive-edge consensus is introduced in [43]. Recently, the positivity-preserving consensus problem of homogeneous MASs is investigated in [52]. Additionally, the problem of formation control for the group of positive interconnected systems and time delays is studied in [53] and [54]. Furthermore, for homogeneous positive systems, stability criteria and consensus availability for these systems are investigated in [55] and [56]. Finally, the corresponding consensus conditions for directed and linear positive MASs with control input nonlinearity are proposed in [57] and then, the non-negative global consensus conditions are proposed. Another interesting application of positive MASs is in providing smart control strategies for distributed energy management networks [58, 59] or investigating smart homes' flexible energy systems [60]. Besides, since in complex networks, using pinning control only requires applying controllers on a small fraction of the nodes, the pinning control strategy has an advantage over other techniques. It is worth noting that pinning consensus has not been proposed for linear or nonlinear positive MASs.

Although some recent practical methods have been developed for analyzing directed positive MASs, other new procedures should be introduced and designed to cope with different consensus protocols of linear and nonlinear positive systems. This matter encourages us to assume the non-negative pinning consensus problem of nonlinear positive MASs with directed graphs and nonlinear dynamics, like uncertainties in modeling and sector input nonlinearities.

Inspired by the reviewed approaches, we present the global leader-following non-negative consensus of nonlinear directed positive MASs with sector input nonlinearities by defining the full-order and distributed observer-based protocols. The contribution of the paper is characterized by the following four aspects. First, a novel distributed pinning-based observer is proposed for positive systems to find every agent's state details by sharing their communication network. Second, nonlinear positive MASs are generally considered and therefore, as the particular case of the current paper to develop any containment tracking/formation control solutions of linear MASs with different types of observer-based protocol, the presented methods of the current paper can be applied to any linearized positive MASs around their equilibrium points. For instance, if there are no isolated and pinned nodes in the network and the agents' dynamics are linearized around their equilibrium points. Then, this article's obtained results can be applied to develop containment formation control for linear systems like the one developed in [61] as long as the linear MAS states remain positive. It is worth noting that most approaches on pinning consensus assume pinning a fraction of nodes. However, a few studies for conventional MASs and none for linear or nonlinear positive MASs consider pinning only one node. Third, the consensus is proven for the case that only one node is assumed to be pinned. Fourth, to implement the proposed approach, two multi-step algorithms are provided to calculate the feedback matrices and design the observer-based consensus strategies.

The remainder of the paper is organized as follows. In Section 2, the main problem is given and formulated. Thereafter, non-negative consensus results and conditions with main algorithms and theorems for fully connected directed nonlinear positive MASs with sector input nonlinearities are proposed based on various types of observer-based control schemes. Two numerical examples are discussed in Section 3 to illustrate the formulated theoretical effectiveness of the outcomes.

*Notations and preliminaries*: $R^{m\times m}$ $\left(R_+^{m\times m}\right)$ signifies the $m\times m$ dimensional matrix with real elements ( non-negative matrix with real elements) set. An $m\times m$ dimensional matrix $Q$ is determined to be Metzler if its non-diagonal elements are considered to be non-negative. Moreover, the matrix $Q(Q\geq 0)$ denotes a non-negative matrix if its elements are completely non-negative. The matrix $Q$ is determined to be positive if all of its elements are non-negative. Furthermore, $Q>0$ denotes a positive definite matrix. As a result, $Q<0$ denotes a negative definite matrix. In formulations, $\lambda_i(Q)$, ( $i=1,\ \ldots,\ N$) is the $i$-th largest eigenvalue of the matrix $Q$. $\|\bullet\|$ shows the Euclidean norm of a vector and the induced norm of a matrix. The $m\times m$ dimensional identity matrix is denoted as $I_m$. Also, the $m\times p$ dimensional zero matrix is shown as $0_{m\times p}$. For simplicity, the symmetrical elements of a symmetric matrix are written with symbol "$*$". In deriving equations, if the dimensions of matrices are not directly stated, the dimensions are considered to be applicable for the algebraic calculations. Let a directed graph (digraph) $\mathcal{G}$ with $N$ vertices ( $v_1,\ v_2,\ \ldots,\ v_N$) indicates the interaction of the presented positive MAS. Denote $\tilde{A}=\{a_{ij}\}\in R^{N\times N}$ as the adjacency matrix of a digraph $\mathcal{G}$, where $a_{ij}=1$ if there is a directed path from $v_j$ to $v_i$, otherwise $a_{ij}=0$. In addition, the Laplacian matrix of the directed network is stated as $L=\tilde{D}-\tilde{A}$, where $\tilde{D}$ is defined as a diagonal matrix represents the in-degree of vertex $v_i$ with $\tilde{d}_{ii}=\sum_{j\neq i}a_{ij}$. Also, the symbol $Sym$ denotes the symmetric matrix.

## 2. Problem description and main results

Consider $N$ nonlinear positive MASs including sector nonlinearities in control inputs with the $i$-th agent's dynamics as given below

$$\begin{cases} \dot{x}_i\left(t\right)=Ax_i\left(t\right)+Hg\left(t,x_i\left(t\right)\right)+Bsec\left(u_i\left(t\right)\right) \\ y_i\left(t\right)=Cx_i\left(t\right), \qquad\qquad i=1,2,\ ..\ .,\ N, \end{cases} \tag{1}$$

where $x_i\left(t\right)\ \in\ R_+^{m\times 1}$ represents the $i$-th agent's state and $A\ \in\ R^{m\times m}$ is considered to be Metzler. Also, $y_i\left(t\right)\in\ R_+^{q\times 1}$ is the measured non-negative output matrix of the $i$-th agent. Moreover, matrices $B$, $C$, and $H$ are constant with applicable dimensions. The above nonlinear system is assumed to be positive if the matrix $A$ remains Metzler and the nonlinear functions remain positive for any positive initial states [62]. It is assumed that $(A, B, C)$ is observable and controllable. The term $sec\,(u_i\,(t))$ stands for the sector nonlinearity of the $i$-th control input, where $u_i\,(t) \in R^{p\times 1}$ is the corresponding user-designed control law and is denoted as $sec\,(u_i\,(t)) = [\; sec\,(u_{1,i}\,(t)) \quad sec\,(u_{2,i}\,(t)) \;\ldots\; sec\,(u_{p,i}\,(t))\;]$.

**Definition 1.** Generally, the nonlinear function $\Phi(z)$ is assumed to be inside the sector $[m_l\; m_h]$ and

$$m_{l,i}{z_i}^2 \le z_i\Phi_i \le m_{h,i}{z_i}^2, \quad i = 1, 2, ..., m,$$

where $(z_i, \Phi_i)$ is the $i$-th scalar element of $(z, \Phi)$.

Based on the sector nonlinearity's above definition, for the control input $u_i$, we can be conclude that

$$m_l u_{ij}^2(t) \le \; u_{ij}\,(t)\, sec\,(u_{ij}\,(t)) \;\le\; m_h u_{ij}^2(t), \tag{2}$$

where $0 \le m_l \le \; m_h < \; \infty$, $i \,\epsilon\, \{\,1,2,\; \ldots,\; N\}$, $j \,\epsilon\, \{\,1,2,\; \ldots,\; p\}$ and $p, q \;\in\; \{1,2,\; \ldots,\; m\}$. Additionally, $g\,(\cdot,\cdot) \;: R \times R_+^{m\times 1} \rightarrow\; R_+^{m\times 1}$ is a continuous function representing the nonlinear dynamics of the $i$-th agent. Furthermore, it is considered that the nonlinear function $g$ meets the following Lipschitz criterion with a Lipschitz fixed value $\beta$ as

$$\|g\,(t, x(\,t)) - g(t,\; y(t))\| \;\le\; \beta\; \|x(t) - y(t)\|\;\;,\; t > 0.$$

For any initial states $x_i\,(0)\;\;(\;i = 1, 2,\; \ldots,\; N)$, the consensus problem can be solved asymptotically for the set of nonlinear agents by the designed control protocol $u_i\,(t)$ if the following condition satisfies

$$\lim_{t\to\infty} \|x_i\,(t) - \; x_j\,(t)\| \;= 0,$$

where $x_i\,(t), x_j\,(t) \;\in\; R_+^{m\times 1}$ and $i\;,\; j \;\in\; \{1,2,\; \ldots,\; N\}$.

A consensus state $s\,(t) \;\in\; R_+^{m\times 1}$ of the proposed positive MAS (1) needs to be a feasible trajectory of an isolated node complying with the following defined consensus state's derivative

$$\dot{s}\,(t) = As\,(t) + Hg\,(t, s\,(t))\,, \tag{3}$$

with $s\,(0) = \; s_0$ .

**Lemma 1 ([63]).** For the directed graph $\mathcal{G}$, all of the eigenvalues of its Laplacian matrix involve non-negative real parts. Besides, the eigenvalue of the related right eigenvector $\mathbf{1}_N$ of the Laplacian matrix is zero. It can be claimed that for the directed graph (digraph) $\mathcal{G}$ with a directed spanning tree, zero would be the simple eigenvalue of the Laplacian matrix $L$.

**Definition 2 ([64]).** For any strongly connected directed graph $\mathcal{G}$, the generalized algebraic connectivity value is defined as

$$a\,(L) = \min_{x^T f = 0\;,\quad x\neq 0} \frac{x^T \hat{L} x}{x^T \Gamma x}\;, \tag{4}$$

where $f \;=\; [\; f_1,\; \ldots,\; f_N]^T$ indicates the corresponding left eigenvector of the graph Laplacian matrix $L$ to its zero eigenvalues, in which $f_i \;>\; 0$ and $\sum_{i=1}^{N} f_i \;=\; 1$. Moreover, $\hat{L}$ is defined as $\hat{L} \;=\; \frac{(\;\Gamma L +\; L^T\Gamma)}{2}$, where $\Gamma = diag\{f_1,\; \ldots,\; f_N\}$ and $diag$ implies the diagonal elements of the matrix.

**Lemma 2 ([64]).** If it is assumed that the symmetric matrix $\hat{L}$ cannot be reduced anymore with the elements $\hat{L}_{ij} \leq 0$ ( $i \neq j; i, j \epsilon \{1, 2, \ldots, N\}$) and the fact that $\sum_{j=1}^{N} \hat{L}_{ij} = 0$, it consists $\min_{x^T\zeta=0 \ , \quad x\neq 0} \frac{x^T \hat{L} x}{x^T x} = \lambda_2(\hat{L})$ if $\zeta = [\ \zeta_1,\ \zeta_2,\ \ldots,\ \zeta_N]^T$ is defined as the left eigenvector of the matrix $\hat{L}$ relating to its zero eigenvalue.

### *2.1. Full-order state observer consensus protocol for nonlinear positive multi-agent systems with nonlinear control input*

The following full-order state observer is presented for the considered nonlinear positive multi-agent network (1)

$$\begin{cases} \dot{\hat{x}}_i\,(t) = A\hat{x}_i\,(t) + Hg\,(t, \hat{x}_i\,(t)) + Bsec\,(u_i\,(t)) + E\,(\ y_i\,(t) -\ \hat{y}_i\,(t))\,, \\ \hat{y}_i\,(t) = C\hat{x}_i\,(t)\,, \qquad\qquad i = 1, 2,\ \ldots\ ,\ N, \end{cases} \tag{5}$$

where $\hat{x}_i\,(t)$ signifies the estimation of the system's state $x_i\,(t)$, and $\hat{y}_i\,(t)$ implicates the output of the proposed observer system and $E$ is the feedback gain matrix.

To minimize the control cost, the objective information $s\,(t)$ is considered to be shared with a limited number of agents. Then, below distributed pinning consensus strategy according to the observed details from the system's states is presented as

$$u_i\,(t) = \ \varphi K \sum_{j=1}^{N} a_{ij}\,(\hat{x}_i(t) - \hat{x}_j(t)) - \varphi d_i K\,(\hat{x}_i(t) - s(t)) \tag{6}$$

where $\varphi$ and $K$ are the network coupling intensity and the feedback gain matrix, respectively, and $d_i$ represents the $i$-th node's pinning status, where $d_i = 1$ denotes the $i$-th node is pinned, otherwise $d_i = 0$. Moreover, $a_{ij}$ indicates the element of the adjacency matrix of the developed communication topology $\mathcal{G}$.

By applying the control input protocol (6) and according to Eqs (1) and (5), the equations of the system's state and its estimated value for the $i$-th agent for the nonlinear positive MAS with the sector nonlinearity form of control input are

$$\begin{cases} \dot{x}_i = Ax_i + Hg\,(t, x_i\,(t)) + Bsec\left(\varphi K \sum_{j=1}^{N} a_{ij}\,(\hat{x}_i(t) - \hat{x}_j(t)) - \varphi d_i K\,(\hat{x}_i(t) - s(t))\right), \\ \dot{\hat{x}}_i = A\hat{x}_i\,(t) + Hg\,(t, \hat{x}_i\,(t)) + EC\,(\ x_i(t) -\ \hat{x}_i(t)) + Bsec\left(\varphi K \sum_{j=1}^{N} a_{ij}\,(\hat{x}_i(t) - \hat{x}_j(t)) - \varphi d_i K\,(\hat{x}_i(t) - s(t))\right). \end{cases} \tag{7}$$

by applying the Kronecker product, Eq.(7) is rewritten as

$$\begin{cases} \dot{x} = (I_N \otimes A)x + (I_N \otimes H)g\,(t, x) + (I_N \otimes B)sec\,(\varphi\,(\ L \otimes K)\,\hat{x} -\ \varphi\,(\ \mathcal{D} \otimes K)\,(\hat{x} - s)\ )\,, \\ \dot{\hat{x}} = (I_N \otimes (A - EC))\hat{x} +\ (I_N \otimes EC)x +\ (I_N \otimes H)g\,(t, \hat{x}) \\ \quad +(I_N \otimes B)sec\,(\varphi\,(\ L \otimes K)\,\hat{x} - \varphi\,(\ \mathcal{D} \otimes K)\,(\hat{x} - s)\ )\,. \end{cases} \tag{8}$$

where $\mathcal{D} = diag\{d_i\}$ is the general pinning matrix. By defining the $i$-th agent's state estimation error as $e_i\,(t) =\ x_i(t) -\ \hat{x}_i(t)$, based on (7), it can be concluded that

$$\dot{e}_i\,(t) = (\ A - EC)\,e_i\,(t) + H\tilde{g}_1\,(t, x_i(t),\ \hat{x}_i(t))\,, \tag{9}$$

where $\tilde{g}_1\,(t, x_i(t),\ \hat{x}_i(t)) \ = \ \ g\,(t, x_i\,(t)) - g\,(t, \hat{x}_i\,(t))$. By denoting $e\,(t) = [\ e_1^T\,(t)\,,\ e_2^T\,(t)\,,\ \ldots,\ e_N^T\,(t)]^T$,

$\tilde{g}_1(t, x(t),\ \hat{x}(t)) = [\tilde{g}_1^T(t, x_1(t),\ \hat{x}_1(t)), \tilde{g}_1^T(t, x_2(t),\ \hat{x}_2(t)),\ \dots,\ \tilde{g}_1^T(t, x_N(t),\ \hat{x}_N(t))]^T$,and

$$g(t, \hat{x}(t)) = [g(t, \hat{x}_1(t)),\ g(t, \hat{x}_2(t)),\ \dots,\ g(t, \hat{x}_N(t))]^T,$$

the state estimation error can be rewritten as

$$\dot{e}(t) = I_N \otimes (A - EC)\, e + (I_N \otimes H)\tilde{g}_1(t, x,\ \hat{x}). \tag{10}$$

Define $\overline{\Omega} = \left[\overline{\Omega}_1{}^T,\ \overline{\Omega}_2{}^T,\ \dots,\ \overline{\Omega}_N{}^T\right]^T$ with $\overline{\Omega}_i = \left[\hat{x}_i^T(t),\ e_i^T(t)\right]^T$, $\xi(u) = \left[\xi^T(u_1),\ \xi^T(u_2),\ \dots,\ \xi^T(u_N)\right]^T$ and the general sector nonlinearity vector $\sec(u)$ as $\sec(u) = [sec(u_1),\ sec(u_2),\ \dots,\ sec(u_N)]^T$. In further formulations, let $\sec(u) = \left[\frac{m_h + m_l}{2}\right] u + \xi(u)$. Now, Eq.(8) can be rewritten as

$$\begin{aligned} \dot{\overline{\Omega}} &= \left(I_N \otimes Q_1 + \varphi \frac{m_h + m_l}{2} (L - \mathcal{D}) \otimes Q_2\right)\overline{\Omega} + \varphi s \frac{m_h + m_l}{2} \mathcal{D} \otimes Q_2 + G + \left(I_N \otimes \overline{B}\right)\xi(u), \\ u &= \varphi\left[(L - \mathcal{D}) \otimes \overline{K}\right]\overline{\Omega} + \varphi s\left(\mathcal{D} \otimes \overline{K}\right), \end{aligned} \tag{11}$$

where $Q_1 = \begin{bmatrix} A & EC \\ 0_{m\times m} & A - EC \end{bmatrix}$, $Q_2 = \begin{bmatrix} BK & 0_{m\times m} \\ 0_{m\times m} & 0_{m\times m} \end{bmatrix}$, $G = \begin{bmatrix} (I_N \otimes H)g(t, \hat{x}) & 0_{Nm\times Nm} \\ (I_N \otimes H)\tilde{g}_1(t, x,\ \hat{x}) & 0_{Nm\times Nm} \end{bmatrix}$, $\overline{B} = \begin{bmatrix} B \\ 0_{m\times p} \end{bmatrix}$ and $\overline{K} = \begin{bmatrix} K & 0_{p\times m} \end{bmatrix}$.

Let denote $\Omega$ as $\Omega = ((L - \mathcal{D}) \otimes I_{2m})\overline{\Omega}$, then

$$\begin{aligned} \dot{\Omega} &= \left(I_N \otimes Q_1 + \varphi \frac{m_h + m_l}{2} (L - \mathcal{D}) \otimes Q_2\right)\Omega + \varphi s \frac{m_h + m_l}{2} (L - \mathcal{D})\mathcal{D} \otimes Q_2 + (I_N \otimes G) \\ &\quad + \left((L - \mathcal{D}) \otimes \overline{B}\right)\xi(u), \\ u &= \varphi\left[I_N \otimes \overline{K}\right]\Omega + \varphi s \mathcal{D}\left((L - \mathcal{D})^{-1} \otimes \overline{K}\right). \end{aligned} \tag{12}$$

**Lemma 3.** Since it is considered that $(A, C)$ is observable, then the initial value of the matrix $E$ for further calculations can be selected to certify that $(A - EC)$ remains stable and

$$\min_{w\ \in R_+} \sigma_{min}(A - EC - jwI) > \beta \|H\| \tag{13}$$

where $\beta$ and $H$ are the Lipschitz constant and the defined coefficient matrix of the system's nonlinear function, respectively. Besides, based on (13), the concluded below Riccati equation

$$(A - EC)^T P + P(A - EC) + (\beta \|H\|)^2 PP + I_m + \delta I_m = 0 \tag{14}$$

includes a symmetric positive definite solution $P$ for some $\delta > 0$ which satisfies (13) [65].

**Theorem 1.** According to a fully connected directed nonlinear positive multi-agent network $\mathcal{G}$ with $N$ positive agents, the dynamical multi-agent nonlinear system is interpreted by (1) with sector input nonlinearities. Then, there is a full-order state observer such that the non-negative consensus is obtained according to the proposed distributed pinning consensus strategy (6) if the below conditions are met

1. $A - \left(\varphi \frac{m_h + m_l}{2}\right)(l_{ii} - d_i)\,\eta BB^T\Pi^{-1}$, $i = 1, \dots,\ N$, and $(A - EC)$ are Metzler. Besides, $-BB^T\Pi^{-1}$ and

$-EC$ are non-positive.

2. $\begin{bmatrix} (A+H)\Pi + \Pi\left(A^T + H^T\right) - \gamma_i BB^T + \Pi\frac{\Sigma}{f_i}\Pi & (EC+H)\Pi \\ * & Sym(A\Pi - EC\Pi) \end{bmatrix} < 0,$

where $\Pi = V^{-1}$, $\gamma_i = \varphi(sd_i + 1)a(L - \mathcal{D})(m_h + m_l)\eta - \left(\frac{\lambda_i(\Gamma LL^T\Gamma)}{f_i}\right) - \left(\frac{\eta^2(m_h - m_l)^2}{4f_i}\right)$, $i = 1, \ldots, N$, $\lambda_i\left(\Gamma LL^T\Gamma\right)$ represents the $i$-th eigenvalue of $\Gamma LL^T\Gamma$ and $a(L - \mathcal{D})$ is calculated based on the definition of general algebraic connectivity of the directed graph $\mathcal{G}$, where $L$ is the Laplacian matrix of the assumed graph and $\mathcal{D} = diag\{d_i\}$ is the general pinning matrix.

**Proof of Theorem 1.** The first equation of (12) is given as

$$\dot{\Omega}_i = Q_1\Omega_i + \varphi\frac{\overline{m}}{2}Q_2\left(\sum_{j=1}^{N} a_{ij}(\Omega_i - \Omega_j) - d_i\Omega_i\right) + \varphi s\frac{\overline{m}}{2}d_iQ_2\left(\sum_{j=1}^{N} a_{ij}(\Omega_i - \Omega_j) - d_i\Omega_i\right) + G_i + \overline{B}(\sum_{j=1}^{N} a_{ij}\left(\xi\left(u_i\right) - \xi\left(u_j\right)\right) - d_i\xi\left(u_i\right) \tag{15}$$

where $\overline{m} = m_h + m_l$. Assume the following positive definite Lyapunov function candidate

$$V(t) = V_1(t) + V_2(t) = e^T(t)\left(I_N \otimes P\right)e(t) + \sum_{i=1}^{N} f_i\Omega_i^T F\Omega_i,$$

where $P$ is the positive definite matrix solution of the obtained Riccati equation (14) and matrix $F = diag\{V, V\} \in R^{2m \times 2m}$ with its element $V$ is positive.

Taking the derivative of $V_1(t)$ leads to

$$\dot{V}_1(t) = 2e^T\left(I_N \otimes P\right)\dot{e} = 2e^T\left(I_N \otimes P\right)\left(I_N \otimes (A - EC)e + (I_N \otimes H)\tilde{g}_1(t, e)\right)$$
$$\leq e^T I_N \otimes \left((A - EC)^T P + P(A - EC) + (\beta\|H\|)^2 PP + I_m\right)e$$

where the given inequality can be applied

$$2e_iPH\tilde{g}_1(t, e_i) \leq 2\|Pe_i\| \cdot \|H\tilde{g}_1(t, e_i)\| \leq 2\beta\|H\| \cdot \|Pe_i\| \cdot \|e_i\| \leq (\beta\|H\|)^2 e_i^T PPe_i + e_i^T e_i.$$

Therefore, for $\dot{V}_1(t)$,

$$\dot{V}_1(t) \leq e^T I_N \otimes \left((A - EC)^T P + P(A - EC) + (\beta\|H\|)^2 PP + I_m\right)e \triangleq e^T I_N \otimes \Sigma e$$
$$< \delta e^T(I_N \otimes I_m)e.$$

On the other side, calculating the derivative of $V_2(t)$ gives

$$\dot{V}_2(t) = 2\sum_{i=1}^{N} f_i\Omega_i^T F\left(Q_1\Omega_i\right) + 2\sum_{i=1}^{N} f_i\Omega_i^T F \times \varphi\frac{\overline{m}}{2}Q_2\left(\sum_{j=1}^{N} a_{ij}(\Omega_i - \Omega_j) - d_i\Omega_i\right)$$
$$+ 2\sum_{i=1}^{N} f_i\Omega_i^T F \times \varphi s\frac{\overline{m}}{2}d_iQ_2\left(\sum_{j=1}^{N} a_{ij}(\Omega_i - \Omega_j) - d_i\Omega_i\right)$$

$$+2\sum_{i=1}^{N} f_i\Omega_i^T F\ (G_i+\ \overline{B}\sum_{j=1}^{N} a_{ij}\left(\xi\left(u_i\right)-\ \xi\left(u_j\right)\right)) = 2\sum_{i=1}^{N} f_i\Omega_i^T(FQ_1\Omega_i)$$
$$+2\sum_{i=1}^{N} f_i\Omega_i^T \times \varphi\frac{\overline{m}}{2}FQ_2\left(\sum_{j=1}^{N} a_{ij}(\Omega_i-\ \Omega_j)-d_i\Omega_i\right)$$
$$+2\sum_{i=1}^{N} f_i\Omega_i^T \times \varphi s\frac{\overline{m}}{2}d_iFQ_2\left(\sum_{j=1}^{N} a_{ij}(\Omega_i-\ \Omega_j)-d_i\Omega_i\right)+2\sum_{i=1}^{N} f_i\Omega_i^T(FG_i)$$
$$+2\sum_{i=1}^{N} f_i\Omega_i^T(F\overline{B}\sum_{j=1}^{N} a_{ij}\left(\xi\left(u_i\right)-\ \xi\left(u_j\right)\right)) = \ 2\Omega^T(\Gamma\ \otimes\ FQ_1+\ \varphi\frac{\overline{m}}{2}\ \Gamma\left(L-\ \mathcal{D}\right)\otimes\ FQ_2)\Omega$$
$$+2\Omega^T(\varphi\frac{\overline{m}}{2}\mathcal{D}\Gamma\left(L-\ \mathcal{D}\right)s\otimes FQ_2+(\Gamma\ \otimes\ FG)+(\Gamma L\otimes\ F\overline{B})\xi\left(u\right))\Omega$$
$$\leq \Omega^T(\ \Gamma\otimes\left(FQ_1+Q_1^TF-\varphi a\left(L-\mathcal{D}\right)\left(\overline{m}\right)\eta F\overline{B}\overline{B}^TF-\varphi\mathcal{D}sa\left(L-\ \mathcal{D}\right)\left(\overline{m}\right)\eta F\overline{B}\overline{B}^TF\right))\Omega$$
$$+2\Omega^T\Gamma\ \otimes(\ FG+G^TF)+\ 2\Omega^T(\ \Gamma L\otimes F\overline{B})\xi\left(u\right), \tag{16}$$

and $\eta$ is a positive constant. Because of the feature of sector nonlinearities that $(\sec\left(u\right)\ -m_hu)^T\left(\sec\left(u\right)\ -m_lu\right)\leq 0$, then it is not difficult to conclude that $\xi^T\left(u\right)\xi\left(u\right)\ \leq\left(\frac{(m_h-m_l)^2}{4}\right)u^Tu$. So,

$$\dot{V}_2\ \leq\Omega^T\left(\ \Gamma\otimes\left(FQ_1+Q_1^TF-\varphi a\left(L-\ \mathcal{D}\right)\left(\overline{m}\right)\eta F\overline{B}\overline{B}^TF-\varphi\mathcal{D}sa\left(L-\ \mathcal{D}\right)\left(\overline{m}\right)\eta F\overline{B}\overline{B}^TF\right)\right)\Omega$$
$$+2\Omega^T\Gamma\ \left(FG+G^TF\right)+\ \Omega^T\left(\ \Gamma LL^T\Gamma\otimes F\overline{B}\overline{B}^TF\right)\Omega+\xi^T\left(u\right)\xi\left(u\right)$$
$$\leq\Omega^T\left(\ \Gamma\otimes\left(FQ_1+Q_1^TF-\varphi a\left(L-\ \mathcal{D}\right)\left(\overline{m}\right)\eta F\overline{B}\overline{B}^TF-\varphi\mathcal{D}sa\left(L-\ \mathcal{D}\right)\left(\overline{m}\right)\eta F\overline{B}\overline{B}^TF\right)\right)\Omega$$
$$+2\Omega^T\Gamma\ \left(FG+G^TF\right)+\ \Omega^T\left(\left(\Gamma LL^T\Gamma+\frac{\eta^2\left(m_h-m_l\right)^2}{4}I_N\right)\otimes F\overline{B}\overline{B}^TF\right)\Omega$$
$$=\Omega^T\left(\Gamma\otimes\left(FQ_1+Q_1^TF\right)\right)\Omega-\ \Omega^T\left(\left(\varphi a\left(L-\ \mathcal{D}\right)\left(\overline{m}\right)\eta\Gamma\right)\otimes F\overline{B}\overline{B}^TF\right)\Omega$$
$$-\ \Omega^T\left(\left(\varphi\mathcal{D}sa\left(L-\ \mathcal{D}\right)\left(\overline{m}\right)\eta\Gamma\right)\otimes F\overline{B}\overline{B}^TF\right)\Omega+2\Omega^T\Gamma\left(FG+G^TF\right)$$
$$+\ \Omega^T\left(\left(\Gamma LL^T\Gamma+\frac{\eta^2\left(m_h-m_l\right)^2}{4}I_N\right)\otimes F\overline{B}\overline{B}^TF\right)\Omega=\overline{\overline{\Omega}}^T\left(\Gamma\otimes\left(FQ_1+Q_1^TF\right)\right)\overline{\overline{\Omega}}$$
$$-\overline{\overline{\Omega}}^T\left(\left(\varphi a\left(L-\ \mathcal{D}\right)\left(\overline{m}\right)\eta\Gamma\right)\otimes F\overline{B}\overline{B}^TF\right)\overline{\overline{\Omega}}-\ \overline{\overline{\Omega}}^T\left(\left(\varphi\mathcal{D}sa\left(L-\ \mathcal{D}\right)\left(\overline{m}\right)\eta\Gamma\right)\otimes F\overline{B}\overline{B}^TF\right)\overline{\overline{\Omega}}$$
$$+2\overline{\overline{\Omega}}^T\Gamma\ \left(FG+G^TF\right)+\ \overline{\overline{\Omega}}^T\left(\left(\Xi+\frac{\eta^2\left(m_h-m_l\right)^2}{4}I_N\right)\otimes F\overline{B}\overline{B}^TF\right)\overline{\overline{\Omega}} \tag{17}$$

where $\overline{\overline{\Omega}}=(Z^T\otimes I)\Omega$ and $Z^T\Gamma LL^T\Gamma Z{=}\Xi{=}\ \mathrm{diag}\{\lambda_i\left(\Gamma LL^T\Gamma\right)\}$, $i=1,\ldots,\ N$ and $Z$ is an orthogonal matrix. Thus, for the $\dot{V}(t)$ we can conclude that

$$\dot{V}\leq\ \delta e^T(I_N\otimes I_m)e+\ \overline{\overline{\Omega}}^T\left(\Gamma\otimes\left(FQ_1+Q_1^TF\right)\right)\overline{\overline{\Omega}}-\overline{\overline{\Omega}}^T\left(\left(\varphi a\left(L-\ \mathcal{D}\right)\left(\overline{m}\right)\eta\Gamma\right)\otimes F\overline{B}\overline{B}^TF\right)\overline{\overline{\Omega}}$$
$$-\ \overline{\overline{\Omega}}^T\left(\left(\varphi\mathcal{D}sa\left(L-\ \mathcal{D}\right)\left(\overline{m}\right)\eta\Gamma\right)\otimes F\overline{B}\overline{B}^TF\right)\overline{\overline{\Omega}}+2\overline{\overline{\Omega}}^T\Gamma\ \left(FG+G^TF\right)$$
$$+\overline{\overline{\Omega}}^T\left(\left(\Xi+\frac{\eta^2\left(m_h-m_l\right)^2}{4}I_N\right)\otimes F\overline{B}\overline{B}^TF\right)\overline{\overline{\Omega}}$$

Due to the nonlinearities in the agents' models and control inputs, the estimated states contain some errors in comparison to real states. It is worth noting that when $\dot{V} < 0$, then $\Sigma < 0$, and $\lim_{t\to\infty} e(t)$ will converge to zero with a very small estimation error, which also infers the asymptotically stable conditions. This estimation error can be ignored for real-life plants with relatively low control performance precisions. Additionally, based on states' initial conditions and the network topology, the set of estimated states will reach to the agreement over time. Furthermore, we can conclude that $\overline{\overline{\Omega}}$ and $\xi(u)$ converge to zero for large values of $t$ ( $t \to \infty$) and the consensus is consequently provided. Since $(A - EC)$ remains Metzler and the state variables remain non-negative in the considered MAS to have the nonlinear system's stability and the fact that we proved the consensus is reached over the time-horizon, the consensus value is guaranteed to be non-negative for the defined nonlinear positive MAS. Subsequently, the observed value from each agent's state is applied to provide the optimal distributed pinning consensus strategy. In the simulation results, both estimation and consensus errors are calculated to verify the mentioned claim. Let define $\Delta$ as follows

$$\Delta = e^T I_N \otimes \Sigma e + f_i\left(F(Q_1+G) + (Q_1+G)^T F\right) - (\varphi(\mathcal{D}s+1)a(L-\mathcal{D})(\overline{m})\eta f_i - \lambda_i\left(\Gamma L L^T \Gamma\right)$$

$$-(\frac{\eta^2(m_h-m_l)^2}{4})F\overline{BB}^T F, i = 1, \ldots, N.$$

From (17), it can be inferred that $\dot{V} < 0$ if $\Delta < 0$ for $i = 1, \ldots, N$; that is

$$\overline{\Delta} = \frac{\Delta}{f_i} = \frac{e^T I_N \otimes \Sigma e}{f_i} + F(Q_1+G) + (Q_1+G)^T F - \left(\varphi(\mathcal{D}s+1)a(L-\mathcal{D})(\overline{m})\eta - \left(\frac{\lambda_i\left(\Gamma L L^T \Gamma\right)}{f_i}\right) - \left(\frac{\eta^2(m_h-m_l)^2}{4f_i}\right)\right)F\overline{BB}^T F < 0. \tag{18}$$

Let define $\gamma_i$ as the following

$$\gamma_i = \varphi(d_i s+1)a(L-\mathcal{D})(\overline{m})\eta - \left(\frac{\lambda_i\left(\Gamma L L^T \Gamma\right)}{f_i}\right) - \left(\frac{\eta^2(m_h-m_l)^2}{4f_i}\right), i = 1, \ldots, N. \tag{19}$$

By substituting $Q_1$ into (18) and using (19), we have the given simplified LMI form as

$$\begin{bmatrix} V(A+H) + \left(A^T + H^T\right)V - \gamma_i V B B^T V + \frac{\Sigma}{f_i} & V(EC+H) \\ * & Sym(VA - VEC) \end{bmatrix} < 0. \tag{20}$$

By pre- and post-multiplying Eq.(19) by $\begin{bmatrix} V^{-1} & \\ & V^{-1} \end{bmatrix}$, the given quadratic matrix inequalities are obtained

$$\begin{bmatrix} (A+H)\Pi + \Pi\left(A^T + H^T\right) - \gamma_i B B^T + \Pi\frac{\Sigma}{f_i}\Pi & (EC+H)\Pi \\ * & Sym(A\Pi - EC\Pi) \end{bmatrix} < 0, \tag{21}$$

where $\Pi = V^{-1}$. Consequently, the consensus condition is proposed above. In the rest, the obligatory requirements to keep the non-negative feature of the whole system are presented. From the augmented system (11), the system matrix $I_N \otimes Q_1 + \varphi\frac{m_h + m_l}{2}(L-\mathcal{D}) \otimes Q_2$ should be Metzler, which leads to $A - \left(\varphi\frac{m_h+m_l}{2}\right)(l_{ii} - d_i)\eta B B^T \Pi^{-1}$, $i = 1, \ldots, N$, and $(A - EC)$ are Metzler. Besides, $-BB^T\Pi^{-1}$ and $-EC$ are non-positive.

In conclusion, the non-negative full-order state observer consensus is attained if the LMI in (21) is stabilized concurrently subject to the non-negative constraints. To have appropriate matrices $\Pi$ and $E$, the convex optimization approach is applied to deal with LMIs depending on the non-negative terms. More information is brought in Algorithm 1.

The proof of Theorem 1 is completed. ■

**Algorithm 1.** According to the fully-connected direct nonlinear positive MAS with sector input nonlinearities, with Laplacian matrix $L$ and the general pinning matrix $\mathcal{D} = diag\{d_i\}$, firstly, $f$ and $\lambda_i\left(\Gamma LL^T\Gamma\right)$ should be calculated. Next, set the constants $m_h$ and $m_l$.

*Step 1*: Select the initial value of the matrix $E$ that satisfies (13).

*Step 2*: Choose a positive definite matrix $P$ that satisfies (14) and then calculate $\Sigma$.

*Step 3*: Choose a positive definite matrix $\Pi$, and solve the corresponding LMIs in 1) with the determined terms of the condition and 2) to have an optimally calculated matrix $E_{opt}$ which can minimize $\alpha$, and

1. $\begin{bmatrix} (A+H)\Pi + \Pi\left(A^T + H^T\right) - \gamma_i BB^T + \Pi\frac{\Sigma}{f_i}\Pi & (EC{+}H)\Pi \\ * & Sym(A\Pi - EC\Pi) \end{bmatrix} < \alpha I_{2m},$
2. $A - \left(\varphi\frac{m_h + m_l}{2}\right)(l_{ii} - d_i)\,\eta BB^T\Pi^{-1}$, $i = 1, \ldots, N$, and $(A - EC)$ are Metzler. Besides, $-BB^T\Pi^{-1}$ and $-EC$ are non-positive.

Let $\alpha_{opt1}$ be the minimized value of $\alpha$.

*Step 4*: If $\alpha_{opt1} \leq 0$, then $\Pi$ and $E_{opt}$ are the optimal solutions of the described LMIs. Otherwise, if $\alpha_{opt1} > 0$, let $E = E_{opt}$, and find the solution of the LMIs in 1) with the determined terms of the condition and 2) to have an optimally calculated matrix $\Pi_{opt}$ which can minimize $\alpha$, and

1. $\begin{bmatrix} (A+H)\Pi + \Pi\left(A^T + H^T\right) - \gamma_i BB^T + \Pi\frac{\Sigma}{f_i}\Pi & (EC{+}H)\Pi \\ * & Sym(A\Pi - EC\Pi) \end{bmatrix} < \alpha I_{2m},$
2. $A - \left(\varphi\frac{m_h + m_l}{2}\right)(l_{ii} - d_i)\,\eta BB^T\Pi^{-1}$, $i = 1, \ldots, N$, and $(A + EC)$ are Metzler. Besides, $-BB^T\Pi^{-1}$ and $-EC$ are non-positive.

Let $\alpha_{opt2}$ be the further minimized value of $\alpha$.

*Step 5*: If $\alpha_{opt2} \leq 0$, then $\Pi_{opt}$ and $E_{opt}$ are the optimal solutions of the described LMIs. Otherwise, start over from step 3 again.

### *2.2. Distributed full-order state observer consensus protocol for nonlinear positive multi-agent systems with nonlinear control input*

In the following, another form of the full-order state observer consensus protocol is presented to estimate the state information of each nonlinear positive agent with sector nonlinearity. Because of the exclusivity of the communication topology in a distributed network, the output information of adjacent agents can be applied to ascertain the state observer and reduce the computational complexity [66]. For that reason, the given distributed pinning observer is introduced as

$$\begin{cases} \dot{\hat{x}}_i(t) = A\hat{x}_i(t) + Hg(t, \hat{x}_i(t)) + Bsec(u_i(t)) + \mu d_i E\left(\hat{y}_i(t) - y_i(t)\right) \\ \qquad + \mu d_i E\left(\sum_{j=1}^N a_{ij}(\hat{y}_i(t) - \hat{y}_j(t)) - \sum_{j=1}^N a_{ij}(y_i(t) - y_j(t))\right), \\ \hat{y}_i(t) = C\hat{x}_i(t), \qquad i = 1, 2, \ldots, N, \end{cases} \tag{22}$$

where $\hat{x}_i$ denotes the state estimation, $\hat{y}_i$ is the output of the designed observer and $d_i$ is each agent's pinning information where $d_i = 1$ if the $i$-th agent is pinned, otherwise $d_i = 0$. Also, $\mu$ is the network coupling intensity. Besides, the distributed pinning consensus strategy is proposed as

$$u_i(t) = \varphi K \sum_{j=1}^{N} a_{ij}(\hat{x}_i(t) - \hat{x}_j(t)) - d_i \varphi K(\hat{x}_i(t) - s(t)), \tag{23}$$

where $\varphi$ and $K$ are defined as the network coupling intensity and the specified feedback gain matrix, respectively.

With the above protocol (23), the equations of the $i$-th agent's state and its estimated value of the nonlinear positive MAS with sector nonlinearities are obtained as below

$$\begin{cases} \dot{x}_i = Ax_i(t) + Hg(t, x_i(t)) + Bsec\left(\varphi K \sum_{j=1}^{N} a_{ij}(\hat{x}_i(t) - \hat{x}_j(t)) - \varphi d_i K(\hat{x}_i(t) - s(t))\right), \\ \dot{\hat{x}}_i(t) = A\hat{x}_i(t) + Hg(t, \hat{x}_i(t)) + \mu d_i EC(\hat{x}_i(t) - x_i(t)) + \mu \sum_{j=1}^{N} a_{ij} EC(\hat{x}_i(t) - \hat{x}_j(t)) - \mu \sum_{j=1}^{N} a_{ij} EC(x_i - x_j) \\ \quad + Bsec\left(\varphi K \sum_{j=1}^{N} a_{ij}(\hat{x}_i(t) - \hat{x}_j(t)) - \varphi d_i K(\hat{x}_i(t) - s(t))\right). \end{cases} \tag{24}$$

By applying the Kronecker product, Eq.(24) is rewritten as

$$\begin{cases} \dot{x} = (I_N \otimes A)x + (I_N \otimes H)g(t, x) + (I_N \otimes B)sec(\varphi(L \otimes K)\hat{x} - \varphi(\mathcal{D} \otimes K)(\hat{x} - s)), \\ \dot{\hat{x}} = (I_N \otimes A - \mu(L + \mathcal{D}) \otimes EC)\hat{x} + (I_N \otimes H)g(t, \hat{x}) + \mu(L + \mathcal{D}) \otimes ECx \\ \quad + (I_N \otimes B)sec(\varphi(L \otimes K)\hat{x} - \varphi(\mathcal{D} \otimes K)(\hat{x} - s)), \end{cases} \tag{25}$$

where $\mathcal{D} = diag\{d_i\}$ is the general pinning matrix. By defining the $i$-th agent's state estimation error as $e_i(t) = x_i(t) - \hat{x}_i(t)$, based on (24), it can be concluded that

$$\dot{e}_i(t) = (A - \mu d_i EC)e_i(t) + H\tilde{g}_1(t, x_i(t), \hat{x}_i(t)) + \mu \sum_{j=1}^{N} a_{ij} EC(e_j(t) - e_i(t)), \tag{26}$$

where $\tilde{g}_1(t, x_i(t), \hat{x}_i(t)) = g(t, x_i(t)) - g(t, \hat{x}_i(t))$. By denoting $e(t) = [e_1^T(t), e_2^T(t), \ldots, e_N^T(t)]^T$, the state estimation error can be rewritten as

$$\dot{e}(t) = (I_N \otimes A - \mu(L + \mathcal{D}) \otimes EC)e + (I_N \otimes H)\tilde{g}_1(t, x, \hat{x}). \tag{27}$$

Define $\overline{\Omega} = [\overline{\Omega}_1, \overline{\Omega}_2, \ldots, \overline{\Omega}_N]^T$ with $\overline{\Omega}_i = [\hat{x}_i^T, e_i^T]^T$. Moreover, $\xi(u) = [\xi^T(u_1), \xi^T(u_2), \ldots, \xi^T(u_N)]^T$, and also, the general sector nonlinearity vector is denoted as $\sec(u) = [sec^T(u_1), sec^T(u_2), \ldots, sec^T(u_N)]^T$. In further formulations, let $\sec(u) = \left[\frac{m_h + m_l}{2}\right]u + \xi(u)$.

Now, the Eq.(25) can be rewritten as

$$\begin{aligned} \dot{\overline{\Omega}} &= \left(Q_1 + \varphi\frac{\overline{m}}{2}(L - \mathcal{D}) \otimes Q_2\right)\overline{\Omega} + \varphi s\frac{\overline{m}}{2}\mathcal{D} \otimes Q_2 + G + (I_N \otimes \overline{B})\xi(u) \\ u &= \varphi[(L - \mathcal{D}) \otimes \overline{K}]\overline{\Omega} + \varphi s(\mathcal{D} \otimes \overline{K}), \end{aligned} \tag{28}$$

where $\overline{m} = m_h + m_l$, $Q_1 = \begin{bmatrix} I_N \otimes A & \mu(L + \mathcal{D}) \otimes EC \\ 0_{Nm \times Nm} & I_N \otimes A - \mu(L + \mathcal{D}) \otimes EC \end{bmatrix}$, $Q_2 = \begin{bmatrix} BK & 0_{m \times m} \\ 0_{m \times m} & 0_{m \times m} \end{bmatrix}$,

$$G = \begin{bmatrix} (I_N \otimes H)g(t,\hat{x}) & 0_{Nm\times Nm} \\ (I_N \otimes H)\tilde{g}_1(t,x,\ \hat{x}) & 0_{Nm\times Nm} \end{bmatrix}, \overline{B} = \begin{bmatrix} B \\ 0_{m\times p} \end{bmatrix} \text{ and } \overline{K} = \begin{bmatrix} K & 0_{p\times m} \end{bmatrix}.$$

Let denote $\Omega$ as $\Omega = (\ (L - \ \mathcal{D}) \otimes I_{2m})\,\overline{\Omega}$, then

$$\begin{aligned} \dot{\Omega} &= \left( Q_1 + \ \varphi\frac{\overline{m}}{2}(L - \ \mathcal{D}) \otimes Q_2 \right)\Omega + \ \varphi\frac{\overline{m}}{2}\mathcal{D}(L - \ \mathcal{D})\, s \otimes Q_2 + (I_N \otimes G) + \left((L - \ \mathcal{D}) \otimes \overline{B}\right)\xi(u) \\ u &= \ \varphi\left[\ I_N \otimes \overline{K}\ \right]\Omega + \ \varphi\mathcal{D}\left((L - \ \mathcal{D})^{-1} s \otimes \overline{K}\right). \end{aligned} \tag{29}$$

**Theorem 2.** According to a fully connected directed positive multi-agent network $\mathcal{G}$ with $N$ positive agents, the dynamical multi-agent nonlinear system is interpreted by (1) with sector input nonlinearities. Then, there is a distributed pinning full-order state observer such that the non-negative consensus is obtained according to the proposed fully distributed pinning consensus strategy (23) if the below conditions are met :

1) $A - \left(\varphi\frac{m_h + \ m_l}{2}\right)(l_{ii} - d_i)\,\eta VBB^TV$, $i = 1, \ldots,\ N$ and $I_N \otimes A - \mu(L + \mathcal{D}) \otimes EC$ are Metzler. Besides, terms $-VBB^TV$ and $-\mu(L + \mathcal{D}) \otimes EC$ are non-positive, where $\mu$ is the network coupling intensity.

2) $\begin{bmatrix} \theta & (\mu d_i EC + H)\Pi \\ * & Sym(A\Pi - \mu d_i EC\Pi) \end{bmatrix} < 0,$

where $\theta = (A + H)\,\Pi + \Pi\left(A^T + H^T\right) - \gamma_i BB^T + \Pi\frac{\Sigma}{f_i}\Pi$ , $\Pi = \ V^{-1}$, $\gamma_i = \ \varphi(d_i s + 1)a\,(L - \ \mathcal{D})\,(m_h + \ m_l)\,\eta - \left(\frac{\lambda_i(\Gamma LL^T\Gamma)}{f_i}\right) - \ \left(\frac{\eta^2(m_h - m_l)^2}{4f_i}\right)$, $i = 1, \ldots,\ N$, $\lambda_i\left(\Gamma LL^T\Gamma\right)$ represents the $i$-th eigenvalue of $\Gamma LL^T\Gamma$ and $a\,(L - \ \mathcal{D})$ is calculated based on the definition of general algebraic connectivity of the directed graph $\mathcal{G}$, where $L$ is the Laplacian matrix of the assumed graph $\mathcal{G}$ and $\mathcal{D} = diag\{d_i\}$ is the general pinning matrix.

**Proof of Theorem 2.** By considering the following Lyapunov candidate

$$V(t) = \ V_1(t) + \ V_2(t) = \ e^T(t)\,(\ I_N \otimes P)\,e(t) + \sum_{i=1}^{N} f_i \Omega_i^T F \Omega_i,$$

the procedure of proving this Theorem is as the same as Theorem 1, except in Theorem 2, by substituting the matrix $Q_1$ in Eq.(18) and using Eq.(19), the given new LMI is obtained as

$$\begin{bmatrix} \overline{\theta} & V(\mu d_i EC + H) \\ * & Sym(VA - \mu d_i VEC) \end{bmatrix} < 0. \tag{30}$$

where $\overline{\theta} = V\,(A + H) + \left(A^T + H^T\right)V - \gamma_i VBB^TV + \frac{\Sigma}{f_i}$ . Moreover, the given quadratic matrix inequalities for a distributed full-order state-observer can be rewritten as

$$\begin{bmatrix} \theta & (\mu d_i EC + H)\Pi \\ * & Sym(A\Pi - \mu d_i EC\Pi) \end{bmatrix} < 0, \tag{31}$$

where $\theta = (A + H)\,\Pi + \Pi\left(A^T + H^T\right) - \gamma_i BB^T + \Pi\frac{\Sigma}{f_i}\Pi$ , $\Pi = \ V^{-1}$. Consequently, the consensus condition is proposed above. In the rest, the obligatory terms to keep the states of the system to be non-negative are presented. From the augmented system (28), the system matrix $Q_1 + \varphi\frac{m_h + \ m_l}{2}(L - \ \mathcal{D}) \otimes Q_2$ should be Metzler, which leads to $A - \left(\varphi\frac{m_h + \ m_l}{2}\right)(l_{ii} - d_i)\,\eta VBB^TV$, and $I_N \otimes A - \mu(L + \mathcal{D}) \otimes EC$ , $i = 1, \ldots,\ N$ are Metzler. Besides, $-VBB^TV$ and $-\mu(L + \mathcal{D}) \otimes EC$ are non-positive.

In conclusion, the non-negative distributed full-order state observer consensus is attained if the LMI in (31)

is stabilized concurrently depending on the non-negative constraints. To have appropriate matrices $\Pi$ and $E$, the convex optimization approach is applied to deal with LMIs subject to the non-negative constraints. More information is brought in Algorithm 2.

The proof of Theorem 2 is completed. ■

**Algorithm 2.** According to the fully-connected direct nonlinear positive MAS with sector input nonlinearities, with Laplacian matrix $L$ and the general pinning matrix $\mathcal{D} = diag\{d_i\}$, firstly, $f$ and $\lambda_i\left(\Gamma LL^T\Gamma\right)$ should be calculated. Next, set the constants $m_h$ and $m_l$.

*Step 1*: Select the initial value of the matrix $E$ that satisfies (13).

*Step 2*: Choose a positive definite matrix $P$ that satisfies (14) and then calculate $\Sigma$.

*Step 3*: Choose a positive definite matrix $\Pi$, and find the solution of the corresponding LMIs in 1) with the determined terms of the condition and 2) to have an optimally calculated matrix $E_{opt}$ which can minimize $\alpha$, and

$$1)\begin{bmatrix} \theta & (\mu d_i EC + H)\Pi \\ * & Sym(A\Pi - \mu d_i EC\Pi) \end{bmatrix} < \alpha I_{2m},$$

where $\theta = (A+H)\Pi + \Pi\left(A^T + H^T\right) - \gamma_i BB^T + \Pi\frac{\Sigma}{f_i}\Pi$ , $\Pi = V^{-1}$ 2) $A - \left(\varphi\frac{m_h + m_l}{2}\right)(l_{ii} - d_i)\eta VBB^TV$, and $I_N \otimes A - \mu(L+\mathcal{D}) \otimes EC$ $i = 1, \ldots, N$, are Metzler. Besides, $-VBB^TV$ and $-\mu(L+\mathcal{D}) \otimes EC$ are non-positive.

Let $\alpha_{opt1}$ be the minimized value of $\alpha$.

*Step 4*: If $\alpha_{opt1} \leq 0$, then $\Pi$ and $E_{opt}$ are the feasible solutions of the described LMIs. Otherwise, if $\alpha_{opt1} > 0$, let $E = E_{opt}$, and find the solution of the LMIs in 1) with the determined terms of the condition and 2) to have an optimally calculated matrix $\Pi_{opt}$ which can minimize $\alpha$, and

$$1)\begin{bmatrix} \theta & (\mu d_i EC + H)\Pi \\ * & Sym(A\Pi - \mu d_i EC\Pi) \end{bmatrix} < \alpha I_{2m},$$

where $\theta = (A+H)\Pi + \Pi\left(A^T + H^T\right) - \gamma_i BB^T + \Pi\frac{\Sigma}{f_i}\Pi$ , $\Pi = V^{-1}$

2) $A - \left(\varphi\frac{m_h + m_l}{2}\right)(l_{ii} - d_i)\eta VBB^TV$, and $I_N \otimes A - \mu(L+\mathcal{D}) \otimes EC$ $i = 1, \ldots, N$, are Metzler. Besides, $-VBB^TV$ and $-\mu(L+\mathcal{D}) \otimes EC$ are non-positive.

Let $\alpha_{opt2}$ be the further minimized value of $\alpha$.

*Step 5*: If $\alpha_{opt2} \leq 0$, then $\Pi_{opt}$ and $E_{opt}$ are the feasible solutions of the described LMIs. Otherwise, start over from step 3 again.

**Remark 1.** One of the goals of developing the methods in this article is to derive consensus strategies for positive multi-agent systems in case of pinning only one node to minimize the control cost (optimal global consensus). Furthermore, according to the theorems, any one of the nodes can be pinned and hence, new LMIs will be derived, which will lead to obtaining new optimal feedback matrices. On the other hand, if the number of pinned nodes increases, suboptimal consensus strategies will be obtained in which the control cost is higher than optimal case since controllers are implemented for more than one agent (node).

## 3. Performance evaluation

In this section, two examples are proposed to verify the effectiveness of the derived observer-based consensus protocols for both simple and complex networks. In the first example, the communication topology is assumed to be simple with few number of nodes and in the second example, a communication topology with more nodes is considered. For both simple and complicated cases, the observer-based consensus is calculated according to the presented algorithms in the article. In further calculations, obtained values from each example and algorithm are denoted with subscripts *opt* (as optimal values), $Ex.1$ (as in Example 1), $Ex.2$ (as in Example 2), $Alg.1$ (from Algorithm 1), and $Alg.2$ (from Algorithm 2).

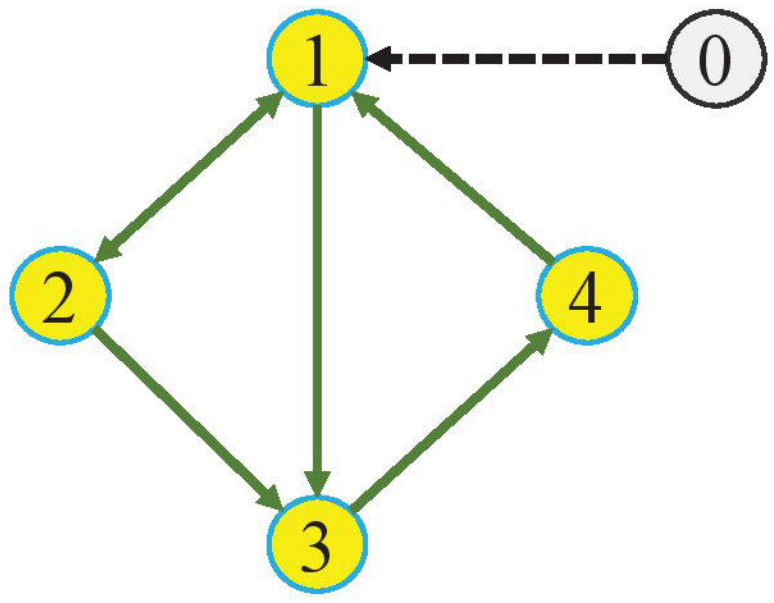

Figure 1: Directed communication topology for nonlinear agents in Example 1 with the isolated objective node( node 0)

**Example 1.** The assumed communication topology of the directed network in the this example is illustrated in Fig.1 with few number of agents. In this case, the eigenvalues of the Laplacian matrix for the considered directed graph are calculated as 3, $1.5 + 0.8660i$ and $1.5 - 0.8660i$.

Let denote parameters of the system(1) as below

$$A = \begin{bmatrix} -10.18 & 9.98 \\ 10.3 & -9.84 \end{bmatrix}, H = \begin{bmatrix} 0.698 & -0.26 \\ 0.682 & -0.34 \end{bmatrix}, B = \begin{bmatrix} 0.7 \\ 0.7 \end{bmatrix}, C = \begin{bmatrix} 0.7 & -0.3 \\ 0.1 & 0.6 \end{bmatrix},$$

and for the $i$-th agent, the control input sector nonlinearity $sec(u_i(t))$ and the nonlinear function $g(t, x_i)$ are defined as follows

$sec(u_i(t)) = 1.2u_i(t) - 0.3u_i(t)\,|\sin(u_i(t))\,|$ and $g(t, x_i) = \begin{bmatrix} x_{i1}\sin(0.31x_{i1}t) & x_{i2}\cos(0.31x_{i2}t) \end{bmatrix}$. In this case, $m_h = 1.2$, $m_l = 0.9$, $\beta = 1$, and $\|H\| = 1.0641$. The system models are chosen to provide the positivity feature of the nonlinear agents before applying the developed algorithms. Then, the algorithms can be applied to preserve the whole system's positivity and calculate the control strategies.Set constants $\eta = 0.6$, $\varphi = 10$, $\mu = 15$, and $\delta = 0.02$. The defined constants in the article are assumed to be known and user-defined values based on the network topology features, including the coupling intensities among agents. By considering various values for the constants which implies the different network structures, new LMIs are solved and new optimal feedback matrices are calculated. In obtaining simulation results, node 1 is considered to be the pinned node.

For the proposed distributed pinning protocol (6), the initial value of the matrix $E$ is chosen as $E = \begin{bmatrix} 17.1333 & 21.8667 \\ 12.3644 & -3.5511 \end{bmatrix}$ based on (13) such that $A - EC$ remains Metzler. Thus, it is easy to obtain the positive definite matrix $P$ from the Riccati equation (14) and then calculate matrix $\Sigma$ as follows.

$P = \begin{bmatrix} 0.0219 & 0.0112 \\ 0.0112 & 0.1357 \end{bmatrix}$, $\Sigma = \begin{bmatrix} -0.0215 & -0.0004 \\ -0.0004 & -0.0198 \end{bmatrix}$. Since there are four agents in the communication network, $N = 4$.

According to the presented procedure in Algorithm 1 and by selecting $\Pi$ as $\Pi = \begin{bmatrix} 0.5 & 0 \\ 0 & 0.5 \end{bmatrix}$, Theorem 1 is solved to calculate local observer values and feedback gain matrix. The following values are extracted from Algorithm 1 and Theorem 1.

$$\alpha_{opt,Ex.1,Alg.1} = -19.026, E_{opt,Ex.1,Alg.1} = \begin{bmatrix} -0.1724 & 3.7982 \\ 3.7982 & 3.0183 \end{bmatrix}, \Pi_{opt,Ex.1,Alg.1} = \begin{bmatrix} 1.05 & 0 \\ 0 & 1.05 \end{bmatrix}.$$

Based on the above matrices, the feedback gain matrix $K$ can be calculated by $K = -\eta B^T\Pi$. So, for Theorem 1, $K_{opt,Alg.1}$ is calculated as

$$K_{opt,Ex.1,Alg.1} = \begin{bmatrix} -0.4410 & -0.4410 \end{bmatrix}.$$

Besides, according to obtained values from Algorithm 1, it is easy to see that $A-\left(\varphi\frac{m_h+m_l}{2}\right)(l_{ii}-d_i)\,\eta BB^T\Pi^{-1}$, $i = 1,\ldots,N$, and $A - E_{opt,Alg.1}C$ are Metzler and $-BB^T\Pi^{-1}$ and $-E_{opt,Alg.1}C$ are non-positive. Therefore, all conditions of Theorem 1 are met, and all obtained values for Theorem 1 are calculated optimally. Then, by Theorem 1, the consensus control problem of the system (1) with distributed pinning control (6) can be asymptotically dealt with. Fig.2 represents the convergence of real states of the system (1), the convergence of observer states, and the calculated estimation error by applying Algorithm 1 and Theorem 1 in setting up the traditional common observer values. Based on Fig.2, we can infer that the proposed local observer by Theorem 1 and Algorithm 1 with a distributed pinning strategy can lead the nonlinear positive MAS with sector input nonlinearities to a non-negative consensus asymptotically.

To obtain distributed full-order state observer as the next proposed observer, the presented LMI in Theorem 2 is solved, and obtained values from Theorem 2 are given as the following.

$$\alpha_{opt,Ex.1,Alg.2} = -12.855, E_{opt,Ex.1,Alg.2} = \begin{bmatrix} -0.0124 & 0.0905 \\ 0.0905 & 0.0784 \end{bmatrix}, \Pi_{opt,Ex.1,Alg.2} = \begin{bmatrix} 1.05 & 0 \\ 0 & 1.05 \end{bmatrix}.$$

thus, based on obtained above values from Algorithm 2 and Theorem 2, the optimal feedback gain matrix is calculated as follows.

$$K_{opt,Ex.1,Alg.2} = \begin{bmatrix} -0.4410 & -0.4410 \end{bmatrix}.$$

additionally, by applying obtained above features from Algorithm 2 and Theorem 2, it can be concluded that $A - \left(\varphi\frac{m_h+m_l}{2}\right)(l_{ii}-d_i)\,\eta BB^T\Pi^{-1}$, and $I_N \otimes A - \mu(L+\mathcal{D}) \otimes E_{opt,Alg.2}C$, $i = 1,\ldots,N$ are Metzler and other conditions of the proposed LMI in Theorem 2 are satisfied. Consequently, by using Theorem 2, the consensus problem of the system (1) with presented distributed full-order state observer in (22), and the distributed pinning strategy (23) can be asymptotically solved. Fig.3 demonstrates the convergence of real and observer states and the estimation error by utilizing Algorithm 2 and Theorem 2 in setting up the distributed full-order state observer values. From Fig.3, we can conclude that the proposed distributed full-order state observer by Theorem 2 and Algorithm 2 with a distributed pinning protocol, the nonlinear positive MAS with sector input nonlinearities can achieve a non-negative consensus asymptotically.

**Example 2.** In this example, a communication topology with more agents is considered to verify the non-negative consensus results for graphs with more complex structure. The communication topology is depicted in Fig.4. The interaction topology in this example is selected from the first case study in [67]. There are nine agents which each agent has two state variables. Thus, for a non-negative consensus, all of eighteen states should converge to the calculated non-negative consensus values. Additionally, the eigenvalues of the

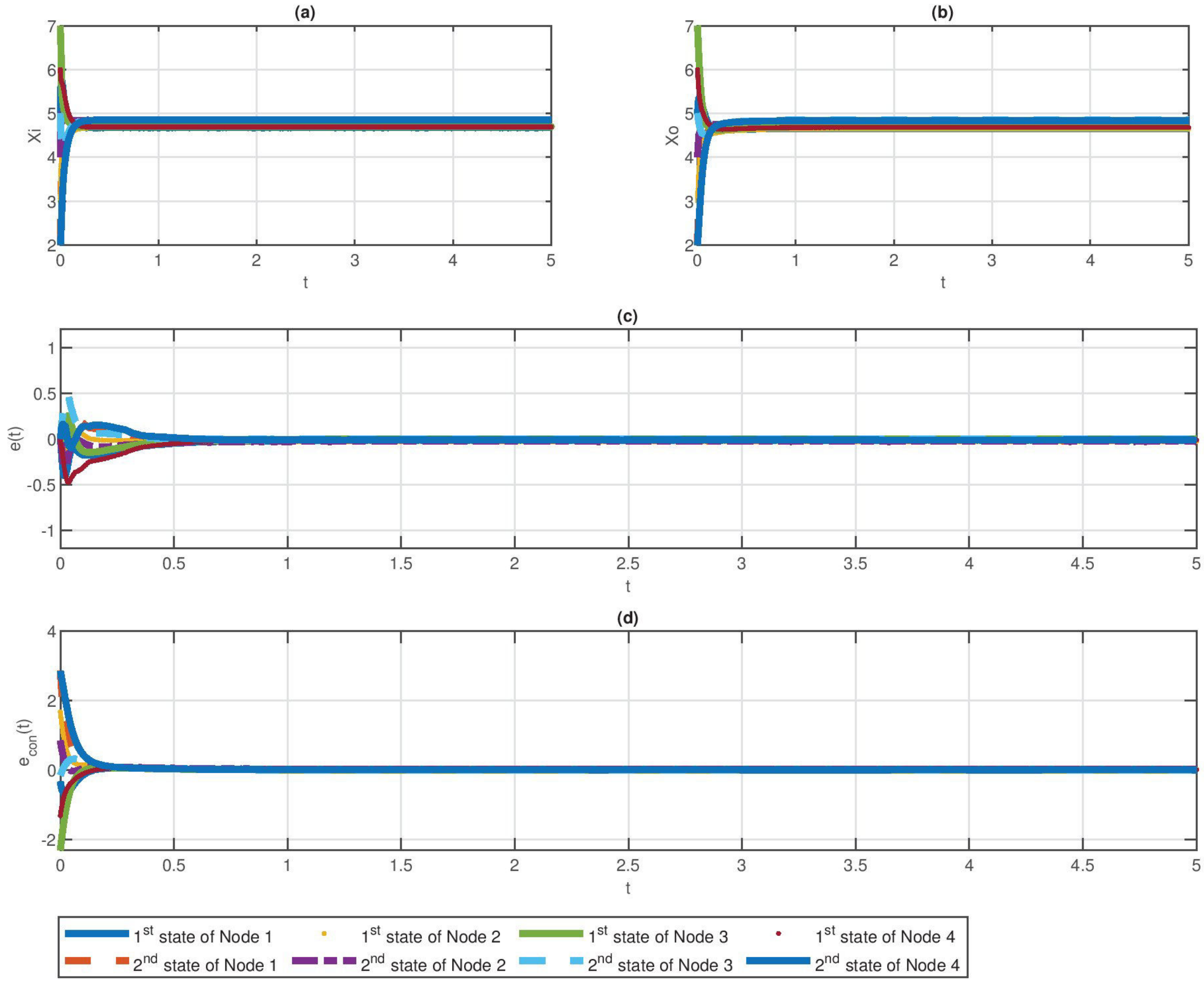

Figure 2: Obtained figures from Example 1: a) evolution of real states, b) evolution of observer states using Algorithm 1, c) estimation error using Algorithm 1, d) consensus error using Algorithm 1; Legends of subplots are the same.

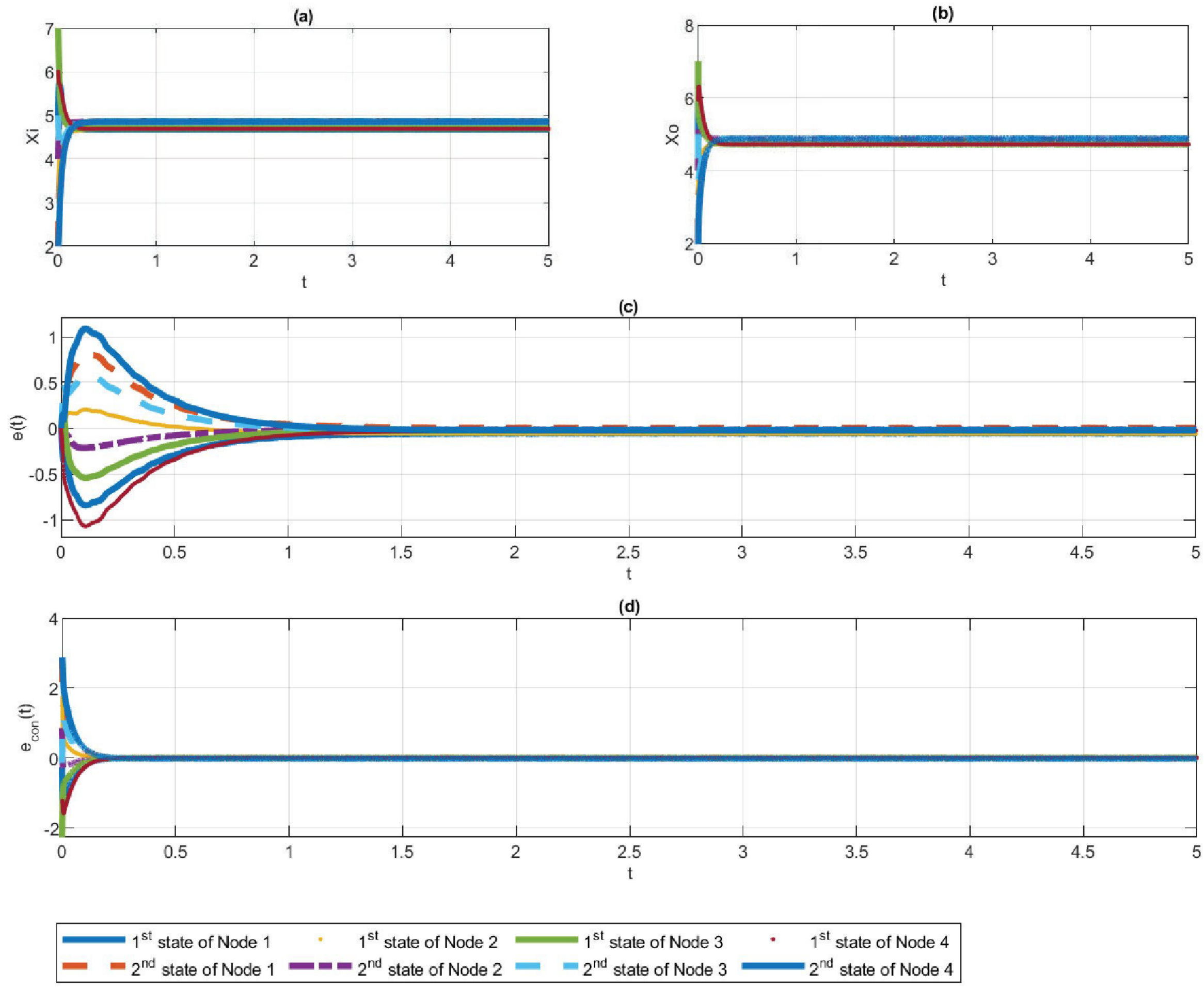

Figure 3: Obtained figures from Example 1: a) evolution of real states, b) evolution of observer states using Algorithm 2, c) estimation error using Algorithm 2, d) consensus error using Algorithm 2; Legends of subplots are the same.

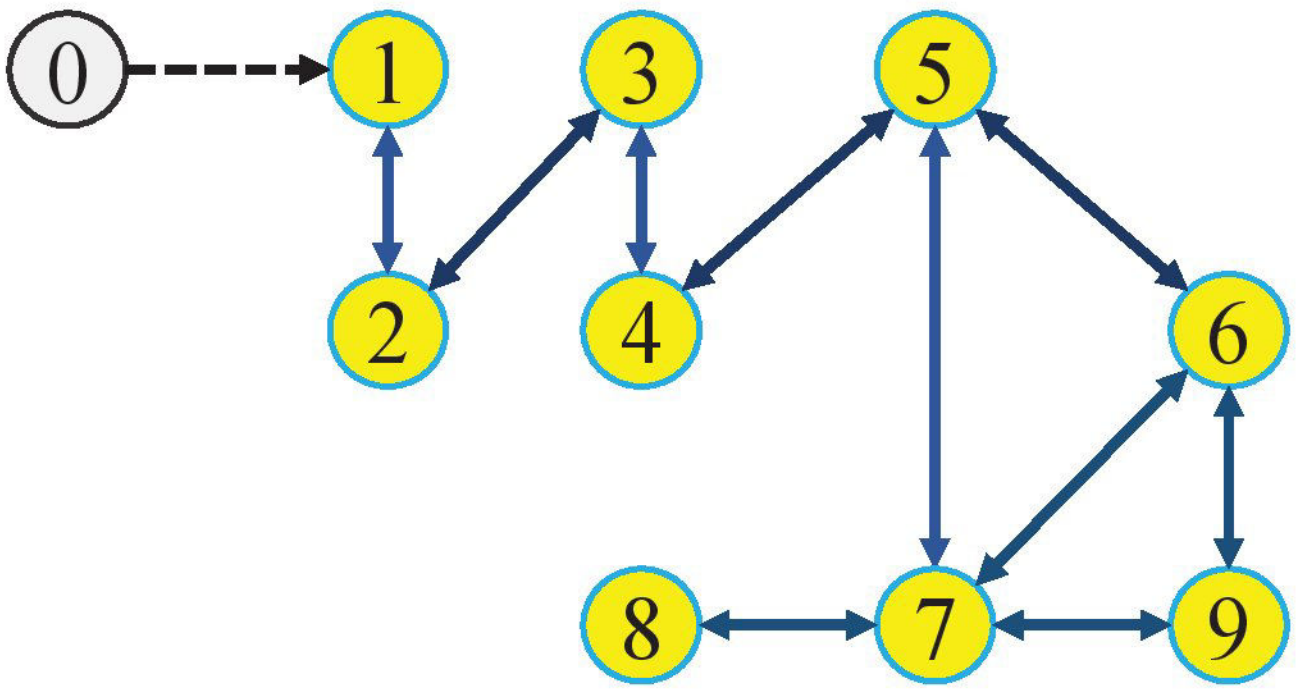

Figure 4: Directed communication topology for nonlinear agents in Example 2 with the isolated objective node( node 0)

Laplacian matrix for the considered directed graph are calculated as 0, 0.1719, 0.8359, 1.1134, 2, 2.8347, 3.5362, 4.3729 and 5.1352.

In this example, the parameters of the system(1) are defined as

$$A = \begin{bmatrix} -13.23 & 8.72 \\ 9.41 & -11.35 \end{bmatrix}, H = \begin{bmatrix} 0.853 & -0.35 \\ 0.841 & -0.44 \end{bmatrix}, B = \begin{bmatrix} 0.82 \\ 0.82 \end{bmatrix}, C = \begin{bmatrix} 0.8 & -0.4 \\ 0.2 & 0.5 \end{bmatrix},$$

and for the $i$-th agent, the control input sector nonlinearity $sec(u_i(t))$ and the nonlinear function $g(t, x_i)$ are defined as the same as in Example.1. Then, $m_h = 1.2$, $m_l = 0.9$, $\beta = 1$, and $\|H\| = 1.3218$. Set constants $\eta = 0.6$, $\varphi = 30$, $\mu = 410$, and $\delta = 0.02$. In obtaining simulation results, node 1 is pinned.

For the proposed distributed pinning protocol (6), the initial value of the matrix $E$ is selected as $E = \begin{bmatrix} 8.7937 & 20.4750 \\ 10.7813 & -6.0750 \end{bmatrix}$ based on (13) such that $A - EC$ remains Metzler. Thus, $P$ and $\Sigma$ are calculated as $P = \begin{bmatrix} 0.0219 & 0.0113 \\ 0.0113 & 0.1357 \end{bmatrix}$, $\Sigma = \begin{bmatrix} -0.0207 & -0.0022 \\ -0.0022 & -0.0080 \end{bmatrix}$. Since there are nine agents in the communication network, $N = 9$.

By applying Algorithm 1 and choosing $\Pi$ as $\Pi = \begin{bmatrix} 0.5 & 0 \\ 0 & 0.5 \end{bmatrix}$, the following optimal values are calculated to satisfy stated conditions in Theorem 1 for the proposed communication topology.

$$\alpha_{opt,Ex.2,Alg.1} = -20.632, E_{opt,Ex.2,Alg.1} = \begin{bmatrix} -0.1345 & 2.3014 \\ 2.3014 & 2.6486 \end{bmatrix}, \Pi_{opt,Ex.2,Alg.1} = \begin{bmatrix} 0.85 & 0 \\ 0 & 0.85 \end{bmatrix}.$$

Based on the above matrices, the optimal feedback gain matrix $K$ is calculated as

$$K_{opt,Ex.2,Alg.1} = \begin{bmatrix} -0.4182 & -0.4182 \end{bmatrix}.$$

Fig.5 illustrates that the asymptotical non-negative consensus is provided based on the calculated optimal values from Theorem 1 for the considered graph.

In the next case, the distributed full-order state observer based on Theorem 2 is presented for the considered communication topology. By applying Algorithm 2, the given optimal values are obtained such that conditions in Theorem 2 are provided.

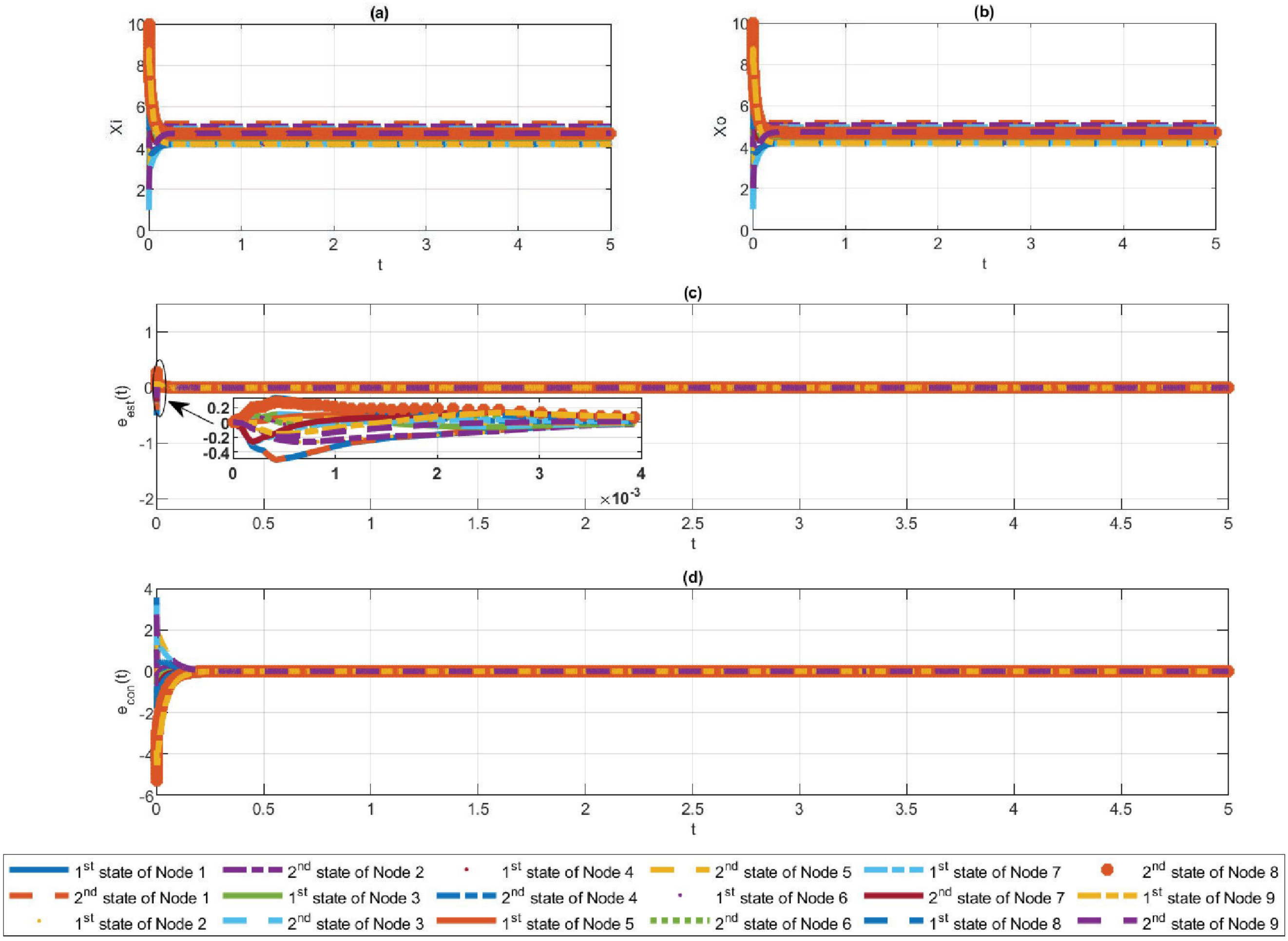

Figure 5: Obtained figures from Example 2: a) evolution of real states, b) evolution of observer states using Algorithm 1, c) estimation error using Algorithm 1, d) consensus error using Algorithm 1; Legends of subplots are the same.

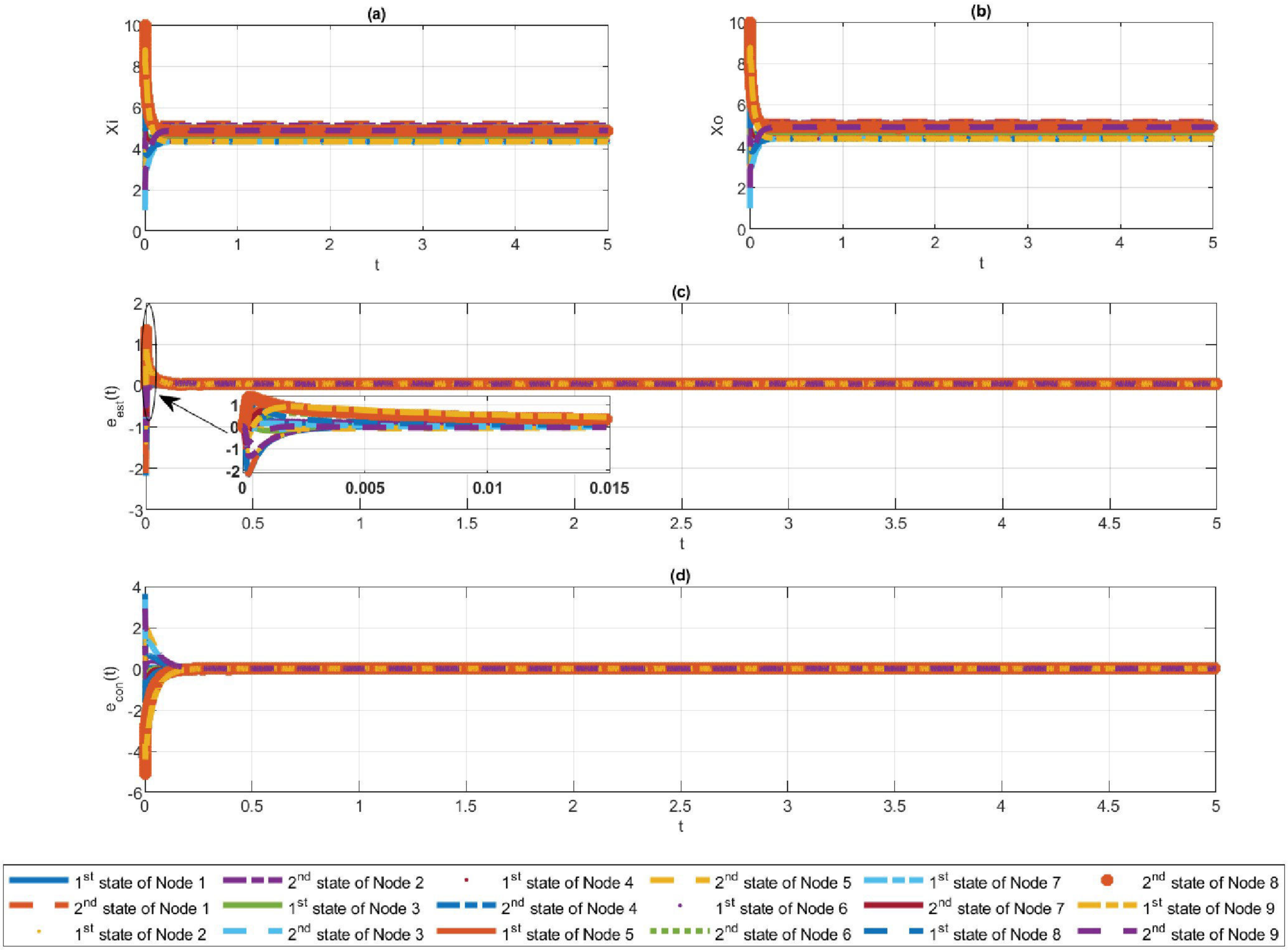

Figure 6: Obtained figures from Example 2: a) evolution of real states, b) evolution of observer states using Algorithm 2, c) estimation error using Algorithm 2, d) consensus error using Algorithm 2; Legends of subplots are the same.

$$\alpha_{opt,Ex.2,Alg.2} = -20.525, E_{opt,Ex.2,Alg.2} = \begin{bmatrix} -0.0005 & 0.0019 \\ 0.0019 & 0.0041 \end{bmatrix}, \Pi_{opt,Ex.2,Alg.2} = \begin{bmatrix} 1.20 & 0 \\ 0 & 1.20 \end{bmatrix}.$$

Based on the above matrices, the optimal feedback gain matrix $K$ is calculated as

$$K_{opt,Ex.2,Alg.2} = \begin{bmatrix} -0.5904 & -0.5904 \end{bmatrix}.$$

Therefore, by applying above values, the non-negative consensus solution of the system (1) with distributed pinning control (6) and the full-order state observer (22) and the distributed pinning strategy (23) is presented. According to Fig.6, the obtained distributed observer-based non-negative consensus for the assumed communication topology is verified. So, for a complex network with nine agents and eighteen state variables, the nonlinear positive MAS with sector input nonlinearities can reach a non-negative consensus asymptotically according to the proved observer-based protocols by Theorems 1 and 2 with one node being pinned. It is obvious that consensus values are dependent to the network topology and the states' initial values. In the simulation results, for each example, the initial conditions of a set of states are set with different values, which leads to obtain different agreements at the same time for different set of states (group consensus). Generally, according to the calculated agreements, by applying algorithms 1 and 2, consensus strategies related to various sets of states are achieved.

**Remark 2.** The developed observer-based algorithms are feasible for distributed networks of positive agents interacting to each other. The obtained observer-based consensus protocol from Algorithm 1 uses the local output information of agents to construct the observers, while in Algorithm 2, observers are made by exchanging the relative output information of neighboring nodes. Since the consensus strategies in both algorithms are proposed by the assumption of applying pinning control method, the implementation requirement will be having the pinning joint communication network with a spanning tree. It can be inferred from the simulation results that Algorithm 1 has higher precision due to the availability of local output information of agents but requires more computations. On the other hand, Algorithm 2 provides consensus with better volatility with less computation efforts because only the relative output information of neighboring agents (not all agents) is needed to construct the observer-based feedback matrices. It is worth noting that applying Algorithm 2 leads to obtaining estimation errors with higher overshoots and longer settling times.

**Remark 3.** Controlling power management systems or data centers' cooling can be considered as controlling positive systems since the control variables in such cases ( PUE, temperature, ...) are always positive. Thus, from the perspective of the process control community, one of the beneficial applications of the presented article is to maintain the output temperature of the liquid-cooling lines in a fixed range (positive consensus) in liquid-cooling methods for data centers with different environmental conditions. Consequently, by applying our article's method, intelligent liquid-cooling strategies can be deployed. This interesting subject will be discussed more with more specific case studies in future publications in line with this article.

## 4. Conclusion

In this paper, the leader-following consensus problem for the directed nonlinear positive MASs with sector input nonlinearities is proposed and formulated according to two types of observer protocols. At first, the traditional local observer is proposed for the nonlinear agent with a nonlinear form of control input to achieve the non-negative distributed pinning consensus protocol. A new type of distributed pinning observer is presented for the networked nonlinear positive MAS by sharing sensors' communication network. Two convex optimization algorithms are stated for both formulated observer protocols to calculate optimal feedback matrices and observers' parameters by solving the quadratic matrix inequalities subject to non-negative terms. Eventually, the leader-following consensus is reached by the distributed pinning strategy according to two proposed effective observer-based methods with two examples provided in numerical illustrations. Based on numerical simulations, both proposed observers can work appropriately with an acceptable estimation error rate in the existence of nonlinearities in agents and their control inputs. Future work will aim to provide non-negative consensus strategies for district heating and cooling of energy management networks and data centers with communication delays.

## References

[1] H. Su, H. Wu, X. Chen, Observer-based discrete-time nonnegative edge synchronization of networked systems, IEEE Transactions on Neural Networks and Learning Systems 28 (10) (2017) 2446–2455.

[2] R. Olfati-Saber, J. A. Fax, R. M. Murray, Consensus and cooperation in networked multi-agent systems, Proceedings of the IEEE 95 (1) (2007) 215–233.

[3] Y. Li, L. Liu, G. Feng, Robust adaptive output feedback control to a class of non-triangular stochastic nonlinear systems, Automatica 89 (2018) 325–332. doi:10.1016/j.automatica.2017.12.020.

[4] G. Wen, W. Yu, Z. Li, X. Yu, J. Cao, Neuro-adaptive consensus tracking of multiagent systems with a high-dimensional leader, IEEE Transactions on Cybernetics 47 (7) (2017) 1730–1742.

[5] G. Wen, P. Wang, T. Huang, W. Yu, J. Sun, Robust neuro-adaptive containment of multileader multiagent systems with uncertain dynamics, IEEE Transactions on Systems, Man, and Cybernetics: Systems 49 (2) (2019) 406–417.

[6] Y. Li, S. Tong, L. Liu, G. Feng, Adaptive output-feedback control design with prescribed performance for switched nonlinear systems, Automatica 80 (2017) 225–231.

[7] H. Hong, W. Yu, G. Wen, X. Yu, Distributed robust fixed-time consensus for nonlinear and disturbed multiagent systems, IEEE Transactions on Systems, Man, and Cybernetics: Systems 47 (7) (2017) 1464–1473.

[8] Z. Liu, X. Yu, Z. Guan, B. Hu, C. Li, Pulse-modulated intermittent control in consensus of multiagent systems, IEEE Transactions on Systems, Man, and Cybernetics: Systems 47 (5) (2017) 783–793.

[9] Z. Li, G. Wen, Z. Duan, W. Ren, Designing fully distributed consensus protocols for linear multi-agent systems with directed graphs, IEEE Transactions on Automatic Control 60 (4) (2015) 1152–1157.

[10] H. Li, X. Liao, T. Huang, Second-order locally dynamical consensus of multiagent systems with arbitrarily fast switching directed topologies, IEEE Transactions on Systems, Man, and Cybernetics: Systems 43 (6) (2013) 1343–1353.

[11] W. Yu, G. Chen, M. Cao, Some necessary and sufficient conditions for second-order consensus in multi-agent dynamical systems, Automatica 46 (2010) 1089–1095. doi:10.1016/j.automatica.2010.03.006.

[12] R. Olfati-Saber, R. M. Murray, Consensus problems in networks of agents with switching topology and time-delays, IEEE Transactions on Automatic Control 49 (9) (2004) 1520–1533.

[13] Q. Song, F. Liu, J. Cao, W. Yu, $m$-matrix strategies for pinning-controlled leader-following consensus in multiagent systems with nonlinear dynamics, IEEE Transactions on Cybernetics 43 (6) (2013) 1688–1697.

[14] Q. Song, J. Cao, W. Yu, Second-order leader-following consensus of nonlinear multi-agent systems via pinning control, Systems & Control Letters 59 (9) (2010) 553–562.

[15] W. Ni, D. Cheng, Leader-following consensus of multi-agent systems under fixed and switching topologies, Systems & Control Letters 59 (3-4) (2010) 209–217.

[16] Y. Hong, J. Hu, L. Gao, Tracking control for multi-agent consensus with an active leader and variable topology, Automatica 42 (7) (2006) 1177–1182.

[17] A. Bidram, A. Davoudi, F. L. Lewis, Z. Qu, Secondary control of microgrids based on distributed cooperative control of multi-agent systems, IET Generation, Transmission Distribution 7 (8) (2013) 822–831.

[18] H. Xin, Z. Lu, Z. Qu, D. Gan, D. Qi, Cooperative control strategy for multiple photovoltaic generators in distribution networks, IET Control Theory Applications 5 (14) (2011) 1617–1629.

[19] G. Schweiger, H. Nilsson, J. P. Schöggl, W. Birk, A. Posch, Modeling and simulation of large-scale systems: A systematic comparison of modeling paradigms, Applied Mathematics and Computation 365 (2019) 124713. doi:10.1016/j.amc.2019.124713.

[20] H. Su, H. Wu, J. Lam, Positive edge-consensus for nodal networks via output feedback, IEEE Transactions on Automatic Control 64 (3) (2019) 1244–1249.

[21] J. Mao, H. R. Karimi, Z. Xiang, Observer-based adaptive consensus for a class of nonlinear multiagent systems, IEEE Transactions on Systems, Man, and Cybernetics: Systems 49 (9) (2019) 1893–1900.

[22] G. Wen, W. Yu, Y. Xia, X. Yu, J. Hu, Distributed tracking of nonlinear multiagent systems under directed switching topology: An observer-based protocol, IEEE Transactions on Systems, Man, and Cybernetics: Systems 47 (5) (2017) 869–881.

[23] S. Li, H. Na-Ni, G. Xin-Ping, Leader-following formation control of multi-agent networks based on distributed observers, Chinese Physics B 19 (2010) 100202. doi:10.1088/1674-1056/19/10/100202.

[24] Z. Li, Z. Duan, G. Chen, L. Huang, Consensus of multiagent systems and synchronization of complex networks: A unified viewpoint, IEEE Transactions on Circuits and Systems I: Regular Papers 57 (1) (2010) 213–224.

[25] Z. Li, L. Xiangdong, P. Lin, W. Ren, Consensus of linear multi-agent systems with reduced-order observer-based protocols, Systems & Control Letters 60 (2011) 510–516. doi:10.1016/j.sysconle.2011.04.008.

[26] Z. Li, L. Xiangdong, W. Ren, L. Xie, Distributed consensus of linear multi-agent systems with adaptive dynamic protocols, Automatica 49. doi:10.1016/j.automatica.2013.03.015.

[27] G. Wen, Z. Li, Z. Duan, G. Chen, Distributed consensus control for linear multi-agent systems with discontinuous observations, International Journal of Control 86 (1) (2013) 95–106. doi:10.1080/00207179.2012.719637.

[28] L. Gao, B. Xu, J. Li, H. Zhang, Distributed reduced-order observer-based approach to consensus problems for linear multi-agent systems, IET Control Theory Applications 9 (5) (2015) 784–792.

[29] B. Liu, D. J. Hill, Z. Sun, Input-to-state-kl-stability and criteria for a class of hybrid dynamical systems, Applied Mathematics and Computation 326 (2018) 124–140.

[30] M. Long, H. Su, B. Liu, Group controllability of two-time-scale multi-agent networks, Journal of the Franklin Institute 355. doi:10.1016/j.jfranklin.2018.06.006.

[31] S. Tong, K. Sun, S. Sui, Observer-based adaptive fuzzy decentralized optimal control design for strict-feedback nonlinear large-scale systems, IEEE Transactions on Fuzzy Systems 26 (2) (2018) 569–584.

[32] S. Tong, Y. Li, Adaptive fuzzy output feedback control for switched nonlinear systems with unmodeled dynamics, IEEE Transactions on Cybernetics 47 (2) (2017) 295–305.

[33] H. Su, Y. Qiu, L. Wang, Semi-global output consensus of discrete-time multi-agent systems with input saturation and external disturbances, ISA Transactions 67. doi:10.1016/j.isatra.2017.01.004.

[34] W. Yu, G. Chen, M. Cao, Consensus in directed networks of agents with nonlinear dynamics, IEEE Transactions on Automatic Control 56 (6) (2011) 1436–1441.

[35] K. Liu, G. Xie, W. Ren, L. Wang, Consensus for multi-agent systems with inherent nonlinear dynamics under directed topologies, Systems & Control Letters 62 (2013) 152–162. doi:10.1016/j.sysconle.2012.11.003.

[36] Z. Li, W. Ren, X. Liu, M. Fu, Consensus of multi-agent systems with general linear and lipschitz nonlinear dynamics using distributed adaptive protocols, IEEE Transactions on Automatic Control 58 (7) (2013) 1786–1791.

[37] K. Takaba, Synchronization of linear agents with sector-bounded input nonlinearities, in: 2015 15th International Conference on Control, Automation and Systems (ICCAS), IEEE, 2015, pp. 1–6.

[38] T. Liu, Z. Jiang, Distributed output-feedback control of nonlinear multi-agent systems, IEEE Transactions on Automatic Control 58 (11) (2013) 2912–2917.

[39] H. Du, S. Li, S. Ding, Bounded consensus algorithms for multi-agent systems in directed networks, Asian Journal of Control 15 (1) (2013) 282–291.

[40] H. Su, M. Z. Q. Chen, J. Lam, Z. Lin, Semi-global leader-following consensus of linear multi-agent systems with input saturation via low gain feedback, IEEE Transactions on Circuits and Systems I: Regular Papers 60 (7) (2013) 1881–1889.

[41] H. Su, M. Z. Chen, X. Wang, J. Lam, Semiglobal observer-based leader-following consensus with input saturation, IEEE Transactions on Industrial Electronics 61 (6) (2013) 2842–2850.

[42] Z. Peng, D. Wang, H. Zhang, G. Sun, Distributed neural network control for adaptive synchronization of uncertain dynamical multiagent systems, IEEE Transactions on Neural Networks and Learning Systems 25 (8) (2014) 1508–1519.

[43] H. Su, H. Wu, X. Chen, M. Z. Q. Chen, Positive edge consensus of complex networks, IEEE Transactions on Systems, Man, and Cybernetics: Systems 48 (12) (2018) 2242–2250.

[44] L. Farina, S. Rinaldi, Positive linear systems: theory and applications, Vol. 50, John Wiley & Sons, 2011.

[45] Y. Ebihara, D. Peaucelle, D. Arzelier, L 1 gain analysis of linear positive systems and its application, in: 2011 50th IEEE Conference on Decision and Control and European Control Conference, IEEE, 2011, pp. 4029–4034.

[46] Y. Ebihara, D. Peaucelle, D. Arzelier, Analysis and synthesis of interconnected positive systems, IEEE Transactions on Automatic Control 62 (2) (2017) 652–667.

[47] Z. Meng, W. Xia, K. H. Johansson, S. Hirche, Stability of positive switched linear systems: Weak excitation and robustness to time-varying delay, IEEE Transactions on Automatic Control 62 (1) (2017) 399–405.

[48] X. Zhao, Y. Yin, L. Liu, X. Sun, Stability analysis and delay control for switched positive linear systems, IEEE Transactions on Automatic Control 63 (7) (2018) 2184–2190.

[49] M. A. Rami, F. Tadeo, A. Benzaouia, Control of constrained positive discrete systems, in: 2007 American Control Conference, IEEE, 2007, pp. 5851–5856.

[50] J. Shen, J. Lam, Containment control of multi-agent systems with unbounded communication delays, International Journal of Systems Science 47 (9) (2016) 2048–2057. arXiv:https://doi.org/10.1080/00207721.2014.971092, doi:10.1080/00207721.2014.971092.
URL https://doi.org/10.1080/00207721.2014.971092

[51] H. Wu, H. Su, Discrete-time positive edge-consensus for undirected and directed nodal networks, IEEE Transactions on Circuits and Systems II: Express Briefs 65 (2) (2018) 221–225.

[52] J. J. R. Liu, J. Lam, Z. Shu, Positivity-preserving consensus of homogeneous multiagent systems, IEEE Transactions on Automatic Control 65 (6) (2020) 2724–2729. doi:10.1109/TAC.2019.2946205.

[53] Y. Ebihara, D. Peaucelle, D. Arzelier, Steady-state analysis of delay interconnected positive systems and its application to formation control, IET Control Theory Applications 11 (16) (2017) 2783–2792.

[54] Y. Ebihara, Convergence rate analysis of delay interconnected positive systems under formation control, in: 2015 10th Asian Control Conference (ASCC), IEEE, 2015, pp. 1–6.

[55] M. E. Valcher, I. Zorzan, New results on the solution of the positive consensus problem, in: 2016 IEEE 55th Conference on Decision and Control (CDC), IEEE, 2016, pp. 5251–5256.

[56] M. E. Valcher, P. Misra, On the stabilizability and consensus of positive homogeneous multi-agent dynamical systems, IEEE Transactions on Automatic Control 59 (7) (2014) 1936–1941.

[57] H. Wu, H. Su, Observer-based consensus for positive multiagent systems with directed topology and nonlinear control input, IEEE Transactions on Systems, Man, and Cybernetics: Systems 49 (7) (2019) 1459–1469.

[58] J. Cai, D. Kim, R. Jaramillo, J. E. Braun, J. Hu, A general multi-agent control approach for building energy system optimization, Energy and Buildings 127 (2016) 337–351.

[59] M. W. Khan, J. Wang, M. Ma, L. Xiong, P. Li, F. Wu, Optimal energy management and control aspects of distributed microgrid using multi-agent systems, Sustainable Cities and Society 44 (2019) 855–870.

[60] A. S. Gazafroudi, T. Pinto, F. Prieto-Castrillo, J. M. Corchado, O. Abrishambaf, A. Jozi, Z. Vale, Energy flexibility assessment of a multi agent-based smart home energy system, in: 2017 IEEE 17th International Conference on Ubiquitous Wireless Broadband (ICUWB), 2017, pp. 1–7. doi:10.1109/ICUWB.2017.8251008.

[61] X. Gong, Y. Cui, J. Shen, Z. Feng, T. Huang, Necessary and sufficient conditions of formation-containment control of high-order multiagent systems with observer-type protocols, IEEE Transactions on Cybernetics (2020) 1–15doi:10.1109/TCYB.2020.3037133.

[62] T. Kaczorek, K. Borawski, Stability of positive nonlinear systems, in: 2017 22nd International Conference on Methods and Models in Automation and Robotics (MMAR), 2017, pp. 564–569. doi:10.1109/MMAR.2017.8046890.

[63] C. Gosil, G. Royle, Algebraic graph theory (2001).

[64] W. Yu, G. Chen, M. Cao, J. Kurths, Second-order consensus for multiagent systems with directed topologies and nonlinear dynamics, IEEE Transactions on Systems, Man, and Cybernetics, Part B (Cybernetics) 40 (3) (2010) 881–891.

[65] R. Rajamani, Observers for lipschitz nonlinear systems, IEEE transactions on Automatic Control 43 (3) (1998) 397–401.

[66] J. Hu, J. Cao, J. Yu, T. Hayat, Consensus of nonlinear multi-agent systems with observer-based protocols, Systems & Control Letters 72 (2014) 71–79. doi:https://doi.org/10.1016/j.sysconle.2014.07.004. URL https://www.sciencedirect.com/science/article/pii/S0167691114001546

[67] L. Hao, X. Zhan, J. Wu, T. Han, H. Yan, Fixed-time group consensus of nonlinear multi-agent systems via pinning control, International Journal of Control, Automation and Systems 19 (1) (2021) 200–208.